\documentclass{aa}  

\usepackage{graphicx}
\usepackage{txfonts}
\usepackage{lipsum}
\usepackage{subcaption}         
\usepackage{lscape}             
\usepackage{placeins}           
                                
\usepackage[colorlinks=true,allcolors=blue,breaklinks=true]{hyperref}
\usepackage{natbib}
\usepackage{booktabs}
\usepackage{soul}
\usepackage{amsmath}
\usepackage{calrsfs}
\usepackage{physics}
\usepackage{upgreek}
\usepackage{multirow}

\DeclareMathAlphabet{\pazocal}{OMS}{zplm}{m}{n}%
\newcommand{\XMM}{\textit{XMM-Newton}}
\newcommand{\Chandra}{\textit{Chandra}}
\newcommand{\swift}{\textit{Swift}}
\newcommand{\La}{\mathcal{L}}%
\newcommand{\LX}{L_\mathrm{X}}
\newcommand{\LS}{L_\mathrm{6\,\upmu m}}
\newcommand{\mnh}{N_\mathrm{H}}
\newcommand{\nh}{$N_\mathrm{H}$}
\newcommand{\dlog}{\mathrm{dlog}}
\newcommand{\pden}{{p_\mathrm{den}}}
\newcommand{\fabs}{f_\mathrm{abs}}
\newcommand{\fctk}{f_{\mathrm{CTK},r}}

\newcommand{\Msun}{M_\sun}
\newcommand{\microns}{\mathrm{\upmu m}}

\begin{document}

    \title{X-ray luminosity function of Compton-thick AGN in the early Universe ($z \geq 3$)}
    \subtitle{Robustness and biases of the CTK population}

    \authorrunning{Ruiz et al.}

    \author{
        A. Ruiz\inst{1}
        \and E. Pouliasis\inst{1}
        \and I. Georgantopoulos\inst{1}        
    }

    \institute{
        IAASARS, National Observatory of Athens, Ioannou Metaxa and Vasileos Pavlou GR-15236, Athens, Greece\\
        \email{ruizca@noa.gr}
    } 
%
%
%



   \date{Received November 7th, 2025  ; Accepted June 8th, 2026 }

  \abstract{The population of Compton-thick (CTK) active galactic nuclei (AGN) represents a critical yet elusive phase in the growth of supermassive black holes (SMBHs). Constraining their abundance and evolution at high redshift is essential for understanding both SMBH growth and the origin of the cosmic X-ray background. We investigate the X-ray luminosity function (XLF) of CTK AGN at $z \geq 3$ using one of the largest available samples of X-ray-selected AGN at high redshift, containing 811 sources from \XMM\ XXL-N and \Chandra\ CCLS and CDF-S/N surveys. We first selected a subsample of ten high-probability CTK candidates, identified through Bayesian X-ray spectral fitting. Their multiwavelength properties are examined through spectral energy distribution modelling to assess the robustness of their CTK classification. For most sources, the inferred X-ray luminosities—and consequently their hydrogen column densities (\nh)—appear overestimated when compared with their infrared (IR) luminosities. After updating the \nh\ posteriors with IR-informed priors, only three sources remain consistent with the CTK regime. To compute the XLF for the entire CTK AGN population, we used $24\,\microns$ photometry to estimate IR luminosities and update the X-ray posteriors for all the remaining sources. Incorporating IR priors systematically reduces the inferred CTK number densities, yielding a more conservative and physically consistent estimate of the X-ray luminosity and absorption functions. We find that CTK AGN constitute $17^{+12}_{-11}\%$ of the total AGN population at $3 \leq z \leq 6$, consistent with results at lower redshifts. Our analysis reveals no statistically significant evolution in the CTK fraction up to $z\sim 6$, suggesting that the most heavily obscured accretion phase remains a persistent component of black hole growth throughout cosmic history. While the overall obscured AGN fraction ($\mnh > 10^{23} \mathrm{cm^{-2}}$) increases toward higher redshifts, the stable CTK fraction supports the interpretation that, at these epochs, the interstellar medium in typical host galaxies cannot produce CTK levels of obscuration.}


   \maketitle
    \nolinenumbers
\section{Introduction}

Active galactic nuclei (AGN) are among the most luminous sources in the Universe, powered by accretion onto supermassive black holes (SMBHs) in their centres \citep{LyndenBell1969}. These black holes have masses typically in the range $10^6-10^9\Msun$. Matter infalling towards the SMBH forms an accretion disc radiating in the extreme ultraviolet (UV), while X-rays are produced by inverse Compton scattering of these photons in a hot corona of electrons above the disc \cite[e.g.][]{Haardt1991}. The primary X-ray emission is often absorbed by gas and dust \citep{Alexander2012,Brandt2015,Netzer2015}, historically interpreted as a toroidal structure \citep{Antonucci1985}, though infrared (IR) and sub-mm imaging reveals a more complex morphology  \citep{Honig2012,GarciaBurillo2016}.

Because X-rays penetrate large column densities, they offer a powerful means of tracing AGN activity across cosmic time. Observations with \Chandra\ and \XMM\ have mapped the accretion history of SMBHs up to $z\sim 3$ \citep[e.g.][]{Ueda2014,Aird2015,Buchner2015,Miyaji2015,Fotopoulou2016a,Georgakakis2017}. At higher redshifts ($z=3-6$), the AGN luminosity function has been explored by \citet{Georgakakis2015}, \citet{Vito2018}, and \citet{Pouliasis2024}, who showed that AGN were more abundant at earlier epochs, with the number density peak depending on luminosity. These studies also established that obscuration increases with redshift: the fraction of obscured AGN with $\log \mnh>23$ rises from $\sim50\%$ at $z \approx 1$ \citep{Signorini2023,Vijarnwannaluk2024} to $\sim75\%$ at $z \geq 3$ \citep{Vito2018,Pouliasis2024}. 
Lower column densities of the order of $\log \mnh\sim 22$ cannot be easily 
constrained at high redshifts as 
the obscuration low energy turnover  is redshifted out of the \XMM\
and \Chandra\ bandpass.
The increase of the column density with redshift  likely reflects the higher gas and dust content of early galaxies, as supported by ALMA observations  \citep{Scoville2017,Gilli2022}.

\begin{table*}[ht]
\caption{Compton-thick candidates selected from the \citet{Pouliasis2024} sample of high-redshift AGN. Quoted uncertainties (or lower limits) correspond to 90\% highest probability density intervals.}
\label{tab:ctksample}
\centering
\begin{tabular}{llcclllc}
\hline\hline
Source & Survey & R.A.       & Dec. & Redshift$^a$ & $\log\LX$$^b$       & $\log\mnh$$^c$ & Prob. CTK \\
\hline
hz414292  & XXL-N & 02:19:01.4 & -05:40:11.10 & 3.24$^\dagger$ & $46.2_{-1.1}^{+0.3}$ & $> 23.5$ & 0.70 \\  
hz434324  & XXL-N & 02:22:16.4 & -04:59:33.26 & 3.16 & $45.8_{-1.1}^{+0.3}$ & $> 23.6$ & 0.84 \\
cdfs101   & CDF-S & 03:32:01.6 & -27:44:02.88 & 3.47 & $44.7_{-0.3}^{+0.3}$ & $> 24.4$ & 0.95 \\
cdfs337   & CDF-S & 03:32:18.8 & -27:51:35.29 & \textbf{3.6600} & $44.7_{-0.3}^{+0.4}$ & $24.16_{-0.33}^{+0.53}$ & 0.93 \\
cdfs657   & CDF-S & 03:32:35.2 & -27:52:15.30 & 3.47 & $44.1_{-0.2}^{+0.5}$ & $24.04_{-0.23}^{+0.85}$ & 0.84 \\
lid\_1278 & CCLS & 09:58:38.2 & +01:35:46.76 & 3.01 & $45.9_{-1.0}^{+0.4}$ & $> 23.6$ & 0.87 \\
cid\_965  & CCLS & 10:00:36.5 & +02:18:28.14 & \textbf{3.1780} & $45.4_{-0.5}^{+0.4}$ & $24.16_{-0.34}^{+1.54}$ & 0.86 \\
lid\_439  & CCLS & 10:01:59.0 & +02:48:02.62 & 2.93 & $45.9_{-1.0}^{+0.3}$ & $24.56_{-0.94}^{+1.37}$ & 0.85 \\
cdfn158   & CDF-N & 12:36:19.9 & +62:19:10.20 & 3.19 & $45.2_{-0.2}^{+0.2}$ & $> 24.5$ & 0.97 \\
cdfn257   & CDF-N & 12:36:36.9 & +62:13:20.10 & \textbf{2.95}$^\dagger$ & $44.4_{-1.3}^{+0.2}$ & $23.5_{-0.3}^{+2.0}$ & 0.72 \\  
\hline
\end{tabular}
\tablefoot{
(a) Bold values correspond to spectroscopic redshifts;
(b) Decimal logarithm of the 2-10~keV luminosity, absorption corrected, in $\mathrm{erg\,s^{-1}}$;
(c) Decimal logarithm of the hydrogen column density, in units of cm$^{-2}$. 
$\dagger$ Updated redshift after visual inspection of the optical and IR counterparts.
}
\end{table*}

For column densities above $\log \mnh \approx 24$, gas becomes optically thick to Compton scattering, defining the Compton-thick (CTK) regime. CTK AGN are key to understanding SMBH growth, AGN–galaxy co-evolution, and the origin of the cosmic X-ray background \citep{Comastri1995,Gilli2007,Akylas2012,Ananna2019}. Hard X-ray (14-195~keV) surveys with \textit{Swift}-BAT \citep{Barthelmy2005} have provided robust constraints in the local Universe ($z \lesssim 0.1$), yielding observed CTK fractions of $\lesssim 20\%$ \citep{Burlon2011,Akylas2016,Georgantopoulos2019,TorresAlba2021} and intrinsic fractions of $\sim20-30\%$ after flux bias corrections \citep{Ricci2015,Akylas2024,Annuar2025}. 

The broad energy range of NuSTAR detectors (8–-79 keV) makes it a powerful tool for the study of CTK-AGN albeit its sensitivity limits its ability to study obscured AGN in the high redshift ($z > 3$) regime. Recent studies based on {\it NuSTAR} observations yield similar results for the fraction of CTK AGN in the local Universe \citep{Boorman2025,Georgantopoulos2025}.  \citet{Ananna2019} reported a fraction of CTK-AGN in the local universe ($z\approx0.1$) that can reach up $50\pm9$\% 
 provided that extremely obscured sources with $\log \mnh>25$ are taken into account. This result is derived from a population synthesis model fitting the observed Cosmic X-ray Background in the 0.5--100~keV energy range.

Beyond the local Universe, CTK AGN have been systematically identified up to about $z=3 $ using \Chandra\ and \XMM\ \citep[e.g.][]{Georgantopoulos2013, Brightman2014,Buchner2015,Lanzuisi2015,Lanzuisi2018,Georgakakis2017,Laloux2023}, but their fraction and evolution remain uncertain. \citet{Buchner2015} reported a CTK fraction of $43\pm10\%$ with no clear redshift dependence, whereas \citet{Lanzuisi2018} found evidence for strong evolution, reaching $\sim45\%$ at $z > 2$. More recently, \citet{Laloux2023} introduced a method combining X-ray spectra with IR priors, enabling the rejection of spurious CTK classifications. Their results are consistent with the local BAT estimates and with no significant evolution of the CTK fraction.

In this work, we revisit the CTK AGN luminosity function at $z > 3$ using the \XMM\ XXL-N, \Chandra\ COSMOS Legacy, and \Chandra\ Deep Field (South and North) surveys. Building on \citet{Pouliasis2024}, who derived the X-ray luminosity and absorption functions (XLAF) for $z = 3-6$, we focus here on the CTK regime ($\mnh \geq 10^{24}~\mathrm{cm}^{-2}$). We re-evaluate CTK candidates through multiwavelength analysis and compare their intrinsic X-ray and IR luminosities to assess the reliability of their CTK classification. These effects are taken into account to calculate the luminosity function for CTK AGN in a more conservative and physically consistent way.

The paper is organised as follows: Sect. 2 describes the initial high-redshift sample and CTK candidate selection; Sect. 3 presents the analysis and results; Sect. 4 discusses our findings; and Sect. 5 summarises our conclusions. Throughout this paper, we adopt a $\Lambda$CDM cosmology with $\mathrm{H_0=70\, km\, s^{-1}\,Mpc^{-1}}$, $\Omega_\mathrm{m}=0.3$ and $\Omega_\Lambda=0.7$ \citep{Komatsu2009}.

\section{Data and sample selection}
\label{sec:samples}

In \citet{Pouliasis2024} we presented one of the largest samples of high-redshift ($z \geq 3$), X-ray-selected (0.5-2~keV band) AGN, comprising 811 sources. The sample was assembled using data from three major X-ray surveys: \Chandra\ Deep Field South and North \citep[CDF-S/N;][]{Luo2017,Xue2016}, \Chandra~COSMOS Legacy Survey  \citep[CCLS;][]{Civano2016}, and the northern field of the \XMM\ XXL survey \citep[XXL-N;][]{Pierre2016a}. The combination of survey depths and sky areas provides broad coverage in X-ray luminosity, redshift, and absorption column density.

The high optical identification rates of these surveys allow the construction of a highly complete sample in terms of distance information, based either on spectroscopic redshifts, when available, or probabilistic photometric redshifts. For the \Chandra\ surveys, \citet{Pouliasis2024} adopted photometric redshifts from the literature \citep{Vito2018,Marchesi2016a}, while for XXL-N, new photometric redshifts were derived using deeper optical imaging from the Hyper Suprime-Cam (HSC) survey \citep{Miyazaki2018}. The final \citet{Pouliasis2024} sample includes all sources with spectroscopic redshifts $z \ge 3$ (191 sources) and those with photometric redshifts having a probability greater than 20\% of being at $z \ge 3$ (620 sources). When weighting the photometric-redshift sources by their high-redshift probabilities, the effective total number of sources is 631.2.

X-ray spectra for all sources were extracted using the standard tools for the corresponding missions, CIAO (v4.13) for \Chandra\ and SAS (v19) for \XMM \citep[See Sect. 3.1 of][for details]{Pouliasis2024} and modelled using the Bayesian X-ray Analysis  framework \citep{Buchner2014,Buchner2021} in combination with the \texttt{UXClumpy} torus model \citep{Buchner2019} to estimate intrinsic X-ray properties \citep[See Sect. 3.2 of][for details on the X-ray spectral modelling]{Pouliasis2024} such as absorption-corrected 2–10 keV luminosities ($\LX$) and hydrogen column densities (\nh).\footnote{Throughout this paper we always use cgs units for $\LX$ ($\mathrm{erg\,s^{-1}}$) and \nh\ ($\mathrm{cm^{-2}}$). When quoting logarithmic values for this quantities, the corresponding cgs units are implied.} For each source, this procedure yields a posterior probability distribution in redshift, luminosity, and absorption, $P(z, \log \LX, \log \mnh | X)$.

\citet{Pouliasis2024} used these posteriors to derive a parametric model for the joint X-ray luminosity and absorption function of AGN (see Appendix~\ref{app:xlaf_details} for a detailed description of the methodology employed) with  $3 \leq z \leq 6$ and $20 \leq \log \mnh \leq 26$, focusing primarily on the unabsorbed and Compton-thin ($22 \leq \log \mnh \leq 24$) regime. In the present work, we extend this analysis to the CTK domain, aiming to assess the reliability of X-ray spectral constraints for potential CTK candidates. To this end, we select a subsample of sources with a high probability of being CTK, based solely on their X-ray properties, as described in Sect.~\ref{sec:ct-sample} below.

\subsection{Sample of CTK AGN}
\label{sec:ct-sample}

We identified compelling candidates for high-redshift CTK AGN within the \citet{Pouliasis2024} sample. Using the posterior probability distributions derived from the X-ray spectral analysis, we computed the marginalised one-dimensional probabilities for $z$ and $\log \mnh$ and selected sources with a probability exceeding 80\% of having $z > 2.5$ and $\log \mnh > 24$. Twelve sources satisfied these criteria. 

\begin{figure}
\centering
\resizebox{\hsize}{!}{\includegraphics{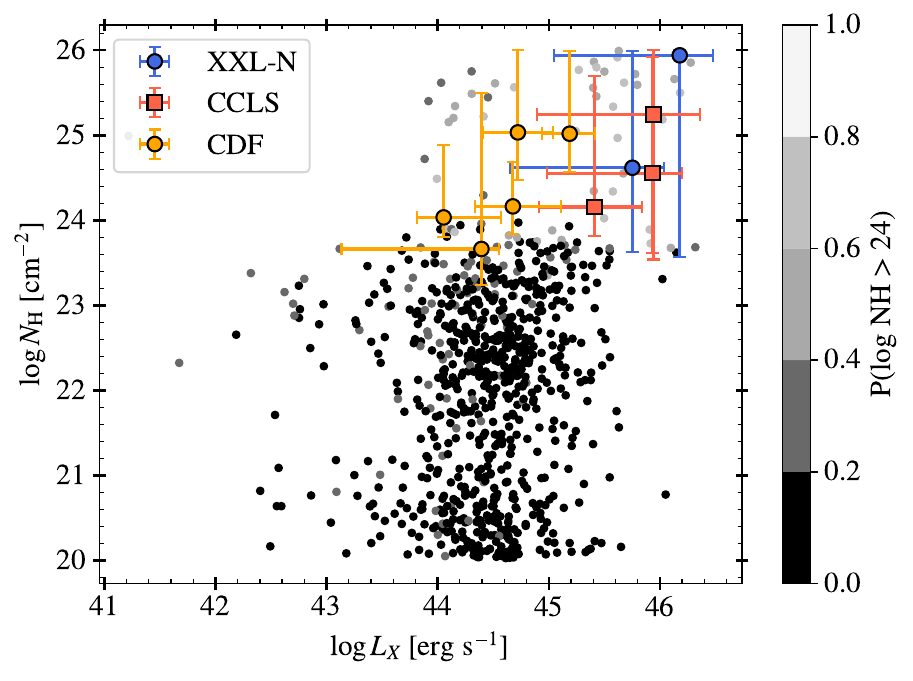}}
\caption{Intrinsic X-ray luminosity in the 2–10~keV band versus hydrogen column density. Small circles show the full sample from  \citet{Pouliasis2024}, shaded by the probability of being CTK. Large circles mark our selection of CTK candidates, as described in Sect.~\ref{sec:ct-sample}, colour-coded according to their parent X-ray survey (yellow: CDF, red: CCLS, blue: XXL-N). For clarity, error bars (indicating 90\% credible intervals) are shown only for the CTK candidates.}
\label{fig:lx_vs_nh}
\end{figure}

We then visually inspected the optical and IR images of these twelve sources to confirm the correct counterparts. Two XXL-N sources (hz428174 and hz414292) were found to have incorrect associations; we recalculated their photometric redshifts using the corrected photometry, after which only hz414292 remained in the sample. One CDF-S source (cdfs412) had an erroneous spectroscopic redshift; its photometric redshift falls below our selection threshold, and it was therefore excluded from the sample. One source in CDF-N (cdfn257) has a JWST spectrum with $z = 2.95$, consistent with our initial photometric redshift estimate. We re-ran the X-ray spectral analysis for all sources with updated redshifts. The final sample comprises ten sources, listed in Table~\ref{tab:ctksample}. Figure~\ref{fig:lx_vs_nh} shows the 2–10~keV luminosity versus the hydrogen column density for the full \citet{Pouliasis2024} sample, highlighting the selected CTK candidates.

\subsection{Multi-wavelength counterparts}
\label{sec:photometry}

We compiled multiwavelength photometric data, spanning from the optical to mid-infrared (MIR) regime, for our sample of CTK AGN using publicly available catalogues from the literature.

For sources in the XXL-N field, we adopted the photometric catalogue of \citet{Pouliasis2024}, which incorporates deeper imaging from recent public data releases and updates the earlier compilation of \citet{Pouliasis2022b}. This catalogue provides optical photometry from HSC \citep[\textit{g, r, i, z, Y} bands;][]{Miyazaki2018}, complemented by \textit{u}-band data from the Canada France Hawaii Telescope \citep[CFHT;][]{Sawicki2019}, and near-infrared (NIR) measurements (\textit{J, H, Ks}) from UKIDSS \citep[DXS/UDS;][]{Lawrence2007} and from several VISTA surveys: VHS \cite[Vista Hemisphere Survey,][]{McMahon2013}, VIKING \citep[VISTA Kilo-degree Infrared Galaxy,][]{Edge2013} and VIDEO \citep[VISTA Deep Extragalactic Observations,][]{Jarvis2013}. We further supplemented this dataset with MIR photometry from the \textit{Spitzer} Enhanced Imaging Products \citep[SEIP;][]{seip} catalogue, including all four IRAC bands (3.6, 4.5, 5.8, and $8.0\,\microns$) and the MIPS $24\,\microns$ band.

For sources in the CCLS field, we used the COSMOS2020 catalogue \citep{Weaver2022}, which combines optical data from HSC (\textit{g, r, i, z, Y}) and CFHT (\textit{u}), NIR data from VISTA (\textit{Y, J, Ks}), and MIR photometry from \textit{Spitzer}–IRAC. We complemented this with \textit{Spitzer}-MIPS $24\,\microns$ fluxes from the HELP catalogues \citep{Shirley2021}.

For the CDF-S field, photometric data were obtained from the ZFOURGE catalogue \citep{Straatman2016}, which includes optical imaging from VLT–VIMOS (\textit{U} band) and HST–ACS (F435W, F606W, F775W, F850LP), complemented by NIR observations from Magellan–FourStar (\textit{J, Hs, Hl}) and MIR data from \textit{Spitzer} (3.6, 4.5, 5.8, 8.0, and $24\,\microns$).

In the CDF-N field, we adopted the photometric compilation of \citet{Xue2016}, who identified optical counterparts to X-ray sources using the HST GOODS-N \citep{Giavalisco2004} and CANDELS \citep{Grogin2011,Koekemoer2011} datasets. This catalogue includes optical and NIR photometry from HST–ACS (F606W) and HST–WFC3 (F125W, F140W, F160W), together with MIR measurements from \textit{Spitzer} covering all IRAC bands and the $24\,\microns$ MIPS channel \citep{Ashby2013}.

\section{Data Analysis}
\label{sec:analysis}

\begin{table*}[t]
\caption{Best-fit model parameters from \texttt{CIGALE} SED fitting. Errors and upper-limits correspond to 1-$\sigma$ confidence intervals.}
\label{tab:sedresults}
\centering
\begin{tabular}{llccccccccc}
\hline\hline
Source $^{(a)}$& Survey & Redshift $^{(b)}$ & red.$\chi^2$ & \multicolumn{2}{c}{AGN fraction} & \multicolumn{2}{c}{Inclination} & \multicolumn{2}{c}{$\log \LS^\mathrm{AGN}$ $^{(c)}$} & $\log\mnh^{IR}$ $^{(d)}$\\
\cmidrule(lr){5-6}\cmidrule(lr){7-8}\cmidrule(lr){9-10}
 & & & & bayes & best & bayes & best & bayes & best \\
\hline
hz414292  & XXL-N & 3.24 & 2.5 & $<0.37$ & 0.1 & $52 \pm 8$ & 50 & $45.3_{-0.5}^{+0.2}$ & 45.2 & $23.7^{+0.7}_{-0.6}$\\
hz434324  & XXL-N & 3.16 & 1.5 & $<0.40$ & 0.2 & $64 \pm 22$ & 30 & $< 44.8$ & 44.0 & $23.5^{+0.5}_{-1.8}$ \\
\textbf{cdfs101}   & CDF-S & 3.47 & 0.5 & $<0.37$ & 0.8 & $63 \pm 18$ & 50 & $44.5_{-0.8}^{+0.3}$ & 44.8 & $24.9\pm1.0$ \\
\textbf{cdfs337}   & CDF-S & \textbf{3.6600} & 3.8 & $0.54 \pm 0.19$ & 0.5 & $75 \pm 14$ & 50 & $45.1_{-0.2}^{+0.1}$ & 44.9  & $24.2\pm0.3$ \\
cdfs657   & CDF-S & 3.47 & 0.8 & $<0.30$ & 0.3 & $71 \pm 19$ & 70 & $< 44.1$ & 43.9 & $24.0\pm0.2$ \\
lid\_1278 & CCLS & 3.01 & 2.0 & $0.51 \pm 0.18$ & 0.7 & $73 \pm 11$ & 70 & $45.0_{-0.2}^{+0.1}$ & 45.2 & $23.3^{+0.4}_{-0.2}$ \\
cid\_965  & CCLS & \textbf{3.1780} & 0.4 & $0.73 \pm 0.10$ & 0.8 & $50 \pm 1$ & 50 & $45.41_{-0.05}^{+0.04}$ & 45.4 & $24.2^{+0.2}_{-0.5}$ \\
lid\_439  & CCLS & 2.93 & 4.8 & $0.63 \pm 0.17$ & 0.5 & $76 \pm 12$ & 70 & $44.7_{-0.2}^{+0.1}$ & 44.8 & $23.6^{+0.13}_{-0.6}$ \\
\textbf{cdfn158}   & CDF-N & 3.19 & 2.4 & $0.35 \pm 0.18$ & 0.3 & $50 \pm 4$ & 50 & $45.5_{-0.2}^{+0.1}$ & 45.5 & $25.0^{+1.0}_{-0.6}$ \\
cdfn257   & CDF-N & \textbf{2.95} & 0.5 & $<0.3$ & 0.4 & $67 \pm 19$ & 50 & $44.1_{-0.8}^{+0.3}$ & 44.2 & $24.5^{+1.5}_{-0.7}$ \\
\hline
\end{tabular}
\tablefoot{
(a) Sources that remain high likelihood CTK ($P(\log\mnh>24)>0.8$ after our SED analysis showed in boldface;
(b) Bold values correspond to spectroscopic redshifts;
(c) Decimal logarithm of the luminosity of the AGN component at $6\,\microns$ in C.G.S. units;
(d) Decimal logarithm of the \nh estimated with the IR-updated posteriors, following the method presented in Sect.~\ref{sec:lxl6}.
}
\end{table*}

\subsection{Spectral energy distributions of the CTK candidates}
\label{sec:seds}

We investigated additional evidence supporting the CTK nature of our selected sources by analysing their multiwavelength properties. CTK AGN are expected to exhibit characteristics typical of Type 2 AGN, where the optical continuum is dominated by the stellar emission of the host galaxy due to the obscuration of the accretion disc by the AGN torus. In contrast, strong thermal emission from the torus is expected in the IR, particularly in the mid-IR, where it can dominate the overall energy output \citep{Nenkova2008a,Nenkova2008b,Sajina2022}.

To explore this, we used the photometric data presented in Sect.~\ref{sec:photometry} to construct the spectral energy distributions (SEDs) of our CTK candidates. These SEDs were modelled using \texttt{CIGALE} \citep[Code Investigating GALaxy Emission;][]{Boquien2019}, a Python-based SED fitting tool designed to disentangle galaxy and AGN emission components and derive their respective physical properties.

In summary, we employed the stellar synthesis population models of \citet{Bruzual2003}, assuming a \citet{Salpeter1955} initial mass function and a fixed metallicity of $Z = 0.02$. The star formation history follows a delayed model of the form $\mathrm{SFR}\propto t\times e^{-t/\tau}$, incorporating a short starburst phase capped at 50 Myr. Dust emission was modelled using the templates of \citet{Dale2014}, excluding any AGN contribution, while dust attenuation was treated using the extinction law of \citet{Charlot2000}. For the AGN component, we adopted the \texttt{SKIRTOR} model \citep{Stalevski2012,Stalevski2016}. The full list of modules and parameters used in our analysis is provided in Table~\ref{tab:cigalesettings}. This setup allowed us to fit the SEDs using a grid comprising over 29 million models.

\begin{figure*}
    \centering
    \includegraphics[width=\textwidth]{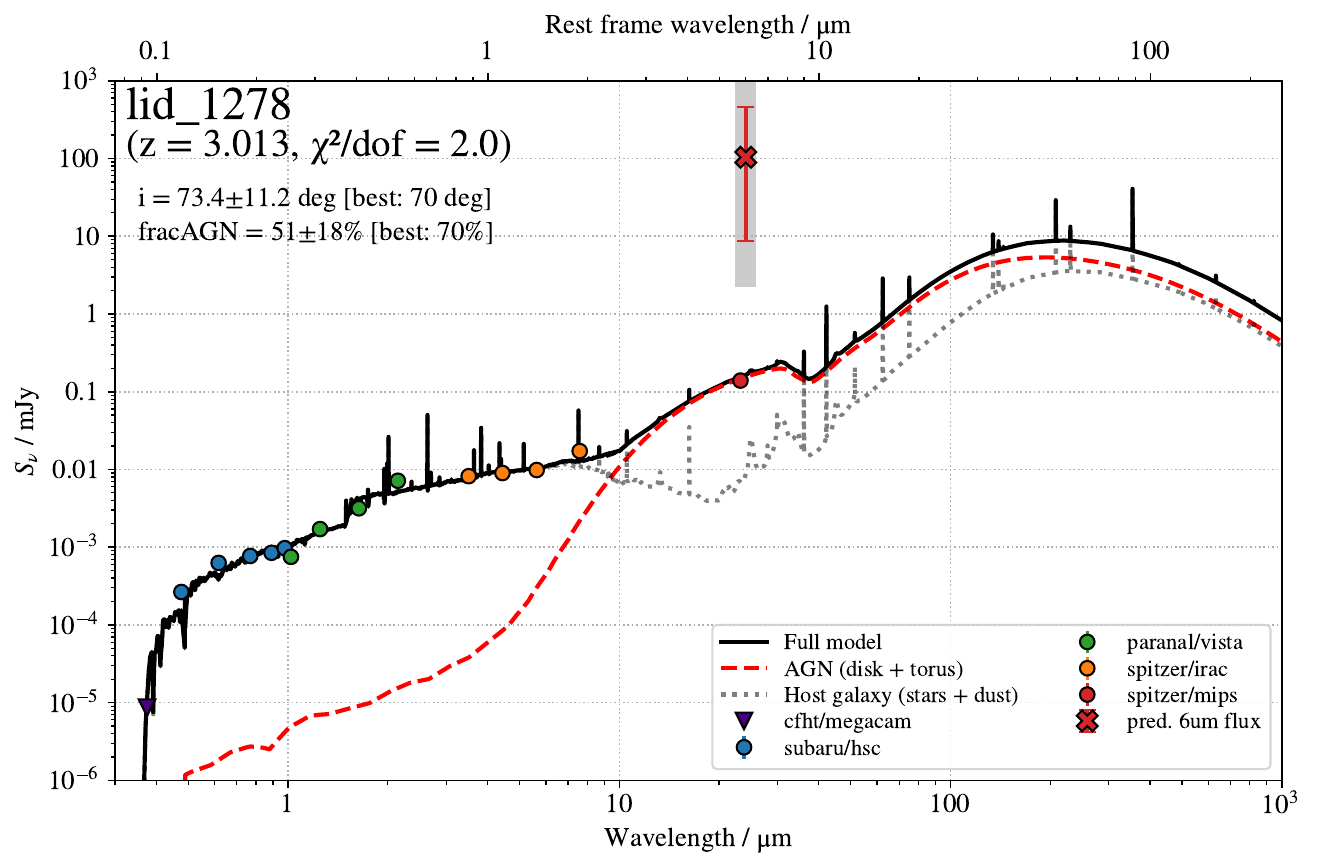}
    \caption{Spectral energy distribution of source lid\_1278 ($z = 3.013$), one of the CTK AGN candidates in our sample. The colour-coded symbols show the observed photometry from CFHT-MegaCam (purple), Subaru-HSC (blue), VISTA (green), \textit{Spitzer}-IRAC (orange), and \textit{Spitzer}-MIPS (red). The solid black line indicates the best-fitting total model from \texttt{CIGALE}, decomposed into the AGN (disc + torus; red dashed line) and host-galaxy (stars + dust; grey dotted line) components. The predicted $6\,\microns$ flux derived from the X-ray luminosity is also shown (red cross). The best-fitting parameters are reported in the legend, including inclination angle and AGN fractional contribution.}
    \label{fig:sedexample}
\end{figure*}

The results of our SED fitting are summarized in Table~\ref{tab:sedresults}. We report three key parameters relevant to our analysis: the inclination angle of the AGN torus, the AGN fractional contribution to the total IR emission, and the AGN luminosity at $6\,\microns$. For each parameter, we provide two estimates: ``best'' refers to the values obtained from the model with the lowest $\chi^2$ in the model grid explored by \texttt{CIGALE}, while ``bayes'' corresponds to the Bayesian estimate (i.e. the probability-weighted mean and standard deviation across all models).\footnote{If the ``bayes'' value is consistent with zero, we quote the $1\sigma$ upper-limit.} As an example, Fig.~\ref{fig:sedexample} shows the best-fit SED model obtained by \texttt{CIGALE} for source lid\_1278.

Our results are consistent with the presence of heavily obscured AGN. The high torus inclination angles indicate that the accretion disc emission is hidden in the optical regime. Although the Bayesian estimates for the AGN fraction are consistent with zero for some sources, visual inspection of the best-fitting SED models clearly shows that an AGN component is required to reproduce the observed  $24\,\microns$ emission. The host-galaxy component alone cannot account for such high infrared fluxes, as also indicated by the non-zero ``best'' AGN fractions reported in Table~\ref{tab:sedresults}. This is clearly illustrated in Fig.~\ref{fig:sedexample}, where the optical part of the SED is dominated by the host galaxy (grey dotted line), whereas the rest-frame $\sim3-30\,\microns$ range is clearly dominated by the AGN component (red dashed line).

Although the overall SED shapes support the CTK scenario, they do not, by themselves, provide strong constraints on the column density of these sources. In Sect.~\ref{sec:lxl6} below, we explore how the AGN luminosity at $6\,\microns$, derived from our SED analysis, can be used to further investigate this aspect.

\subsection{Constraining the X-ray absorption via the $\LX-\LS$ correlation}
\label{sec:lxl6}

A well-established correlation exists between AGN emission in the X-ray and MIR regimes, particularly between the rest-frame 2–10~keV and $6\,\microns$ luminosities \citep[e.g.][]{Mateos2015,Stern2015,Chen2017,Toba2019}. This correlation arises from the shared origin of both emissions: UV photons from the accretion disc serve as seed photons for Compton up-scattering in the X-ray corona, as well as for the thermal reprocessing by the circumnuclear dusty torus, which re-emits in the MIR. Owing to its robustness, this correlation has been used to identify heavily obscured AGN \citep{Alexander2008,Georgantopoulos2011,Severgnini2012,Carroll2021}, and more recently, to place constraints on the line-of-sight \nh\ via X-ray spectral modelling \citep{Laloux2023}.

\begin{figure*}[ht]
    \centering
    \includegraphics[width=0.49\textwidth]{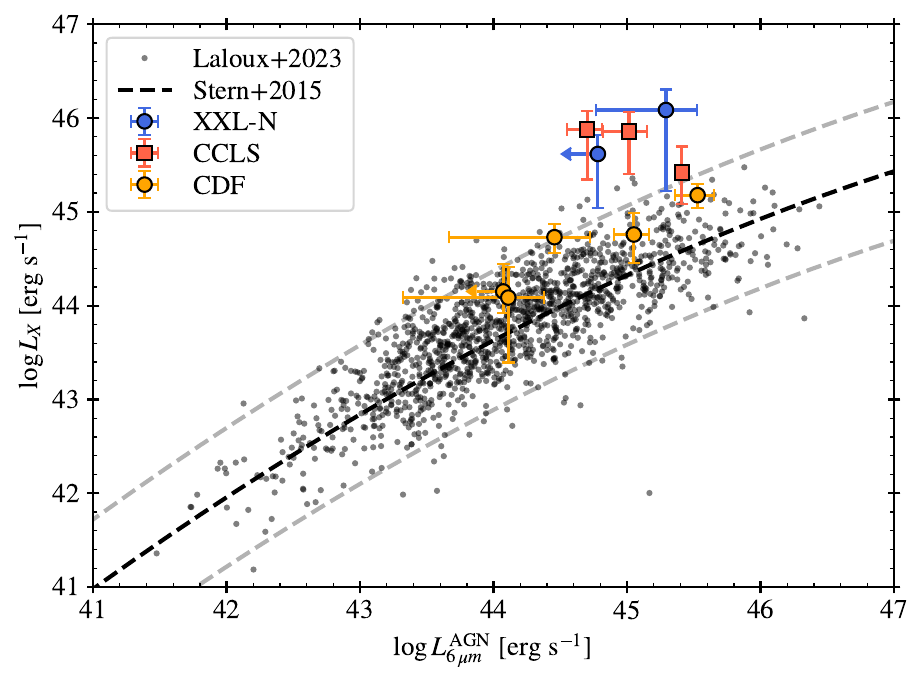}
    \includegraphics[width=0.49\textwidth]{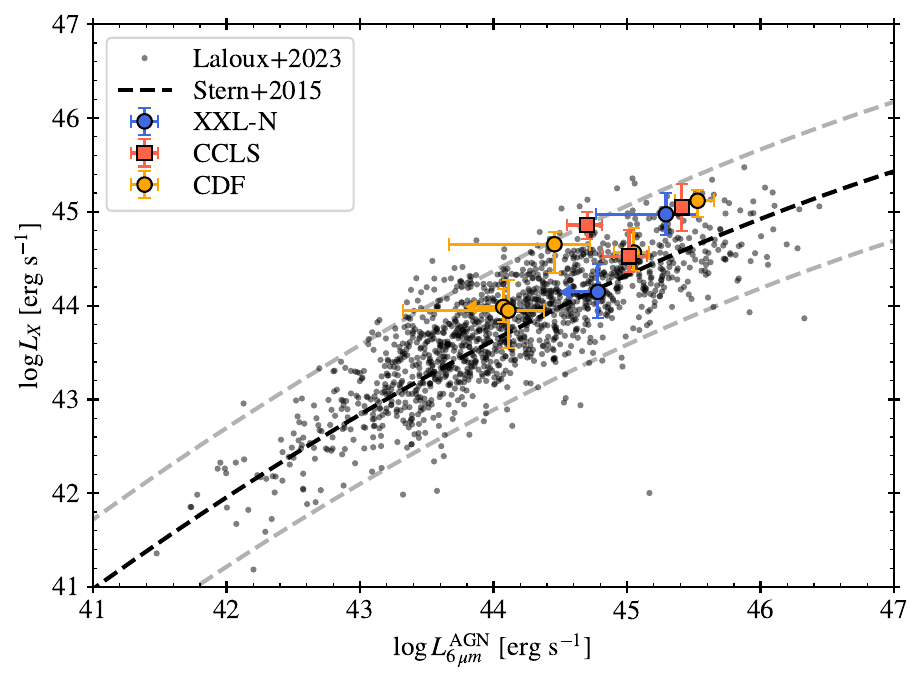}

    \caption{Intrinsic X-ray luminosity in the 2–10 keV band versus the monochromatic $6\,\microns$ luminosity of the AGN component. The results for our CTK sample are shown before (left panel) and after (right panel) applying the infrared luminosity prior, as described in Sect.~\ref{sec:lxl6}. Grey dots represent the X-ray–selected AGN sample from \citet{Laloux2023} in the CCLS field. Large circles indicate our final CTK candidates, colour-coded by their parent X-ray survey (yellow: CDF; red: CCLS; blue: XXL-N). The black dashed line marks the $\LX-\LS$ relation from \citet{Stern2015}, while the grey dashed lines denote the $2\sigma$ dispersion of the \citet{Laloux2023} sample with respect to that relation.}
    \label{fig:lxl6}
\end{figure*}

In the left panel of Fig.~\ref{fig:lxl6}, we present the rest-frame 2–10~keV luminosities derived from our X-ray spectral fits as a function of the rest-frame $6\,\microns$ luminosities obtained through SED modelling (Sect.~\ref{sec:seds}) for the CTK candidates in our sample (coloured circles). For reference, we include a comparison sample of X-ray selected AGN from the CCLS \citep[black dots;][]{Laloux2023}. The distribution of our sources reveals a systematic offset above the \citet{Stern2015} correlation (black dashed line), with the XXL-N and most CCLS sources lying beyond the $2\sigma$ dispersion (gray dashed lines) derived from the \citet{Laloux2023} sample.

This discrepancy is further illustrated in Fig.~\ref{fig:sedexample}, where we show the observed $6\,\microns$ flux inferred from the X-ray luminosities using the \citet{Stern2015} relation (red cross). The expected IR flux is several orders of magnitude higher than what is predicted from our SED modelling, suggesting a significant overestimation of the X-ray luminosity. The availability of \textit{Spitzer}-MIPS $24\,\microns$ imaging data for all CTK candidates provides a strong constraint on the rest-frame $6\,\microns$ emission at the redshift range of our sources, lending confidence to the reliability of the SED-derived IR luminosities.

The most plausible interpretation for the discrepancy is an overestimation of $\LX$, driven by uncertainties in \nh. Inspection of the \nh\ posterior distributions (lower right panels of Figs.~\ref{fig:xrdata_xmm}-\ref{fig:xrdata_chandra}) reveals that they are frequently broad or exhibit bimodal behaviour, with non-negligible posterior weight extending below the CTK threshold.

\begin{figure}[t]
    \centering
    \includegraphics[width=\linewidth,trim={0 0 9cm 0},clip]{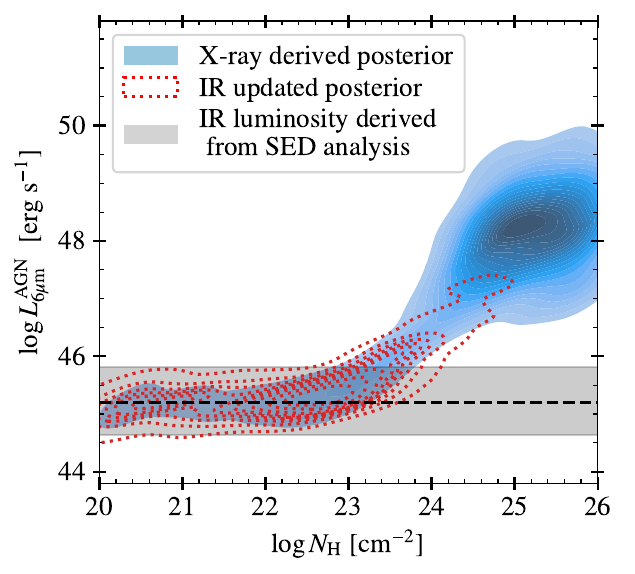}
    \caption{Joint posterior distribution in the $\log\mnh-\log\LS$ plane (blue shaded region) for source lid\_1278, one of our CTK candidates. The infrared luminosity $\LS$ is inferred from the X-ray luminosity posterior via the \citet{Stern2015} relation. The black dashed line and grey shaded region show the SED-inferred $\LS$ value and its uncertainty, while the red dotted contours indicate the posterior distribution after incorporating the IR prior.}
    \label{fig:posteriorexample}
\end{figure}

Figure~\ref{fig:posteriorexample} illustrates the joint posterior distribution in the $\log\mnh-\log\LS$ plane (blue shaded region) for one of our CTK candidates, where $\LS$ is computed from the $\LX$ posterior using the \citet{Stern2015} relation. The distribution is characteristically bimodal, with a primary peak at high \nh\ and high $\LS$, and a secondary peak at lower \nh\ and correspondingly lower luminosity. Notably, the low-\nh\ solution aligns well with the SED-inferred $\LS$ value (black dashed line and gray shaded region), suggesting that the CTK classification may be driven by poorly constrained or degenerate \nh\ estimates rather than robust spectral evidence.

\citet{Laloux2023} demonstrated that the degeneracy in the $\log\mnh-\log\LX$ parameter space, commonly encountered in X-ray spectral fitting in the low-count regime, can be significantly reduced by incorporating IR luminosity constraints. In their approach, IR data are used to inform the X-ray spectral analysis through priors on the intrinsic X-ray luminosity, assuming a given $\LX-\LS$ correlation.

Following the same principle, we incorporate the constraints derived from our SED fitting into the X-ray analysis without re-performing the full spectral fit. Instead, we update the posterior probability distributions obtained with BXA using Bayes' theorem:
\begin{equation}
\begin{split}
    P(z, \log \LX, \log \mnh | X, IR) \propto & P(z, \log \LX, \log \mnh | X) \times \\ 
    & \mathcal{N}(\log \LX|\mu,\sigma^2),    
\end{split}
\end{equation}
where $P(z, \log \LX, \log \mnh | X)$ is the original posterior distribution from the X-ray-only analysis, and 
$\mathcal{N}$ is a Gaussian prior on $\log\LX$ centred at the X-ray luminosity inferred from the SED-derived $\LS$ (assuming the \citealt{Stern2015} correlation). The standard deviation $\sigma$ of the prior is set to match the intrinsic scatter of the $\LX-\LS$ relation measured in the \citet{Laloux2023} sample. The updated posterior, $P(z, \log \LX, \log \mnh | X,IR)$, incorporates the additional IR constraint. This method is illustrated in Fig.~\ref{fig:posteriorexample}, where the red dotted region represents the IR-informed posterior distribution.

\begin{figure}[t]
    \centering
    \resizebox{\hsize}{!}{\includegraphics{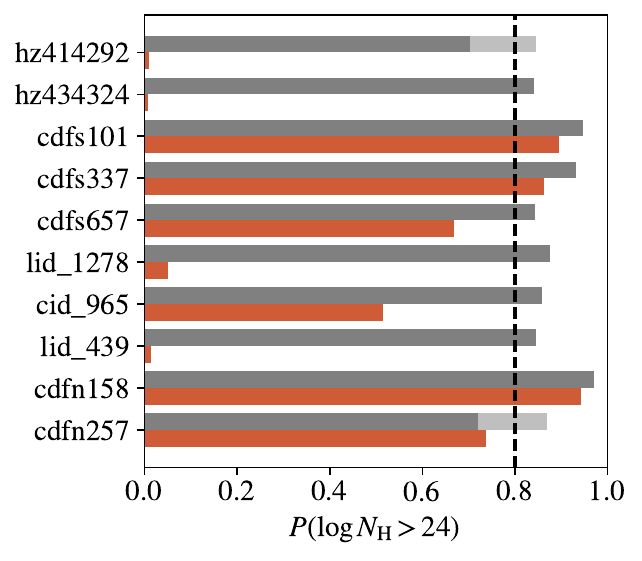}}
    \caption{Probability of the selected CTK candidates of having $\log\mnh > 24$. Grey bars show the original probabilities estimated using the X-ray only posteriors, and red bars show the results using the IR-updated posteriors. Light-grey areas show the estimated probabilities before applying the redshift corrections discussed in Sect.~\ref{sec:ct-sample}. The black dashed vertical line shows our initial selection criterion for the CTK sample.}
    \label{fig:probctk}
\end{figure}

We applied this approach to all CTK candidates in our sample. The right panel of Fig.~\ref{fig:lxl6} compares the updated $\LX$ values against the SED-inferred $6\,\microns$ luminosities. After incorporating the IR priors, the sources fall closer to the expected  relation from \citet{Stern2015}, although a slight systematic overestimation of $\LX$ persists. Importantly, the updated posteriors have a significant impact on the inferred \nh\ values. Figure~\ref{fig:probctk} shows a comparison of the posterior probabilities of each source being CTK, obtained using the X-ray-only (grey bars) and IR-informed (red bars) posteriors. After applying the IR constraint, only three of the original CTK candidates show probabilities above our initial selection threshold of 80\%.

This result highlights the critical importance of including IR data for the reliable identification of CTK sources. Figure~\ref{fig:lx_vs_nh} shows that a significant number of objects in the \citet{Pouliasis2024} sample exhibit nominal \nh\ values above the CTK threshold but were excluded from our analysis due to low posterior probabilities of being CTK. These sources typically display broad \nh\ posterior distributions, and therefore the inclusion of IR constraints could substantially refine their absorption estimates.

For a complete and unbiased characterization of the CTK population—particularly in the context of high-redshift X-ray luminosity function studies (see Sect.~\ref{sec:xlf})—it is essential to obtain IR luminosity estimates for the entire \citet{Pouliasis2024} sample. While a full SED analysis is beyond the scope of this work, we derived approximate $\LS$ values using available MIR imaging.

\begin{figure*}
    \centering
    \includegraphics[width=\linewidth]{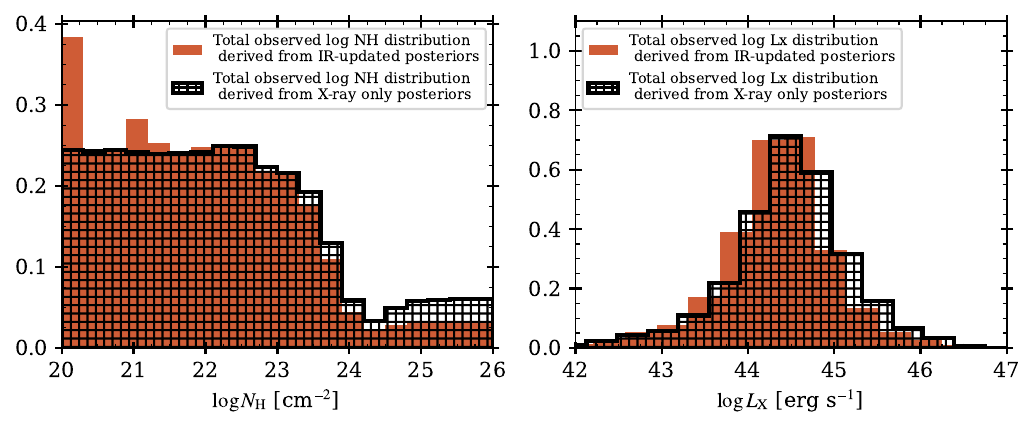}
    \caption{Observed distributions of hydrogen column density and X-ray luminosity for the full \citet{Pouliasis2024} sample. The black-hatched histograms show the original X-ray posteriors, while the red-solid histograms correspond to the posteriors updated with IR information (see Sect.~\ref{sec:lxl6}).}
    \label{fig:lognh_loglx_obsdist}
\end{figure*}

All fields in the sample have been observed by \textit{Spitzer}-MIPS at $24\,\microns$, covering the majority of the survey areas. We retrieved the flux-calibrated MIPS mosaics from the SEIP database\footnote{\url{https://irsa.ipac.caltech.edu/data/SPITZER/Enhanced/SEIP/}} and performed aperture photometry at the optical counterpart positions of all sources in the \citet{Pouliasis2024} catalogue. This yielded observed $24\,\microns$ fluxes for the majority of sources. By adopting a representative AGN template from the \texttt{CIGALE} library, redshifting it appropriately, and scaling to the observed $24\,\microns$ flux, we estimated rest-frame $\LS$ values. Using this method, we derived IR luminosity estimates for 574 out of 811 sources in the \citet{Pouliasis2024} sample.  236 sources in the XXL-N field lie outside the MIPS coverage, so we could not estimate IR luminosities for those sources.

These $\LS$ values should be interpreted as upper limits, since we did not account for potential contamination from host galaxy emission or blending with nearby sources. To incorporate these estimates into the X-ray analysis, we updated the posterior distributions using a conservative prior on $\LX$. Instead of a Gaussian, we used a step function prior: the probability is set to zero for $\log\LX > \log\LX^\mathrm{IR} + 2\sigma$ , and constant otherwise, where $\LX^\mathrm{IR}$ is the X-ray luminosity predicted from $\LS$ using the \citet{Stern2015} relation.

We validated this method by applying it to the ten CTK candidates for which we had SED-derived $\LS$ values, and found consistent results between both methods. We also tested the sensitivity of the outcome to the choice of AGN template, we recalculated the IR luminosities using an alternative AGN template and we verified that the specific template used had minimal impact on the final results, provided it corresponded to a high-inclination (edge-on) configuration.

Fig.~\ref{fig:lognh_loglx_obsdist} shows the impact of the IR-informed priors on the full \citet{Pouliasis2024} sample. The left and right panels display the stacked posterior distributions \citep[Appendix A]{Baronchelli2020} of \nh\ and $\log\LX$, respectively. The black hatched histograms correspond to the original (X-ray-only) posteriors, while the red histograms reflect the IR-updated posteriors. While the overall $\LX$ distribution shows a modest shift toward lower luminosities, the effect on the \nh\ distribution is more pronounced for the CTK population: the observed number density of CTK sources is reduced by approximately a factor of two. The distributions for unabsorbed and Compton-thin sources remain largely unaffected.

\subsection{X-ray luminosity function for CTK}
\label{sec:xlf}

The X-ray luminosity function (XLF) of AGN is a fundamental diagnostic of black hole growth and AGN evolution \citep{Fotopoulou2016a}. By tracing the number density of AGN as a function of luminosity and redshift, the XLF provides a robust observational constraint of accretion activity across cosmic time \citep{Aird2015a}. An additional dimension can be introduced by including the hydrogen column density, \nh, which enables the study of the cosmic evolution of the intrinsic AGN population as a function of X-ray absorption. This leads to the definition of the X-ray luminosity and absorption function (XLAF):
\begin{equation}
\label{eq:xlaf}
    \phi_\mathrm{abs}(\LX,\mnh,z) = \dfrac{\dd[3]{N}(\LX,\mnh,z)}{\dd V \dd\log(\LX) \dd\log(\mnh)} = \phi(\LX,z) \times f_\mathrm{abs}(\mnh,\LX,z),
\end{equation}
where the XLAF, $\phi_\mathrm{abs}(\LX,\mnh,z)$, represents the number of sources $N$ per unit comoving volume $V$, per logarithmic interval in luminosity and column density, as a function of $\LX$, \nh\ and $z$. It is commonly assumed that the XLAF can be factorised into two components \citep[e.g.][]{Ueda2003,Ueda2014,Vijarnwannaluk2022}: the XLF, $\phi(\LX,z)$, and the absorption function, $f_\mathrm{abs}(\mnh,\LX,z)$.

In this work, we follow the methodology of \citet{Pouliasis2024} to derive a parametric form of the XLAF for CTK sources in the redshift range $z = 3-6$. The XLF is parametrised as a broken power law \citep{Barger2005} with pure density evolution \citep[PDE;][]{Schmidt1968}. \citet{Pouliasis2024} tested various evolutionary models for the XLF, finding no statistically significant differences among them when applied to our dataset. We adopt the PDE model here, as it offers a good description of the data with the fewest free parameters.

For the absorption function, we use the parametrisation from \citet{Pouliasis2024}: a flat-step function defined over three \nh\ intervals, with values that depend on both redshift and X-ray luminosity. A detailed description of the XLAF parametrisation is provided in Appendix~\ref{app:xlaf_details}.

As in \citet{Pouliasis2024}, we employ a Bayesian inference framework to estimate simultaneously the parametric forms of the XLF and absorption function. This approach allows a rigorous propagation of the uncertainties in the X-ray spectral parameters and photometric redshifts of individual sources, encoded in their posterior probability distributions. The posterior distribution of the XLAF parameters is sampled using MLFriends \citep{Buchner2016,Buchner2017}, a nested-sampling Monte Carlo algorithm implemented in the \texttt{UltraNest} package. Full details of the methodology are given in Appendix~\ref{app:xlaf_details}.

In addition to the parametric modelling described above, we derived a binned representation of the XLAF using the method of \citet{Miyaji2001}. This approach estimates the space density in each luminosity, \nh\ and redshift bin by scaling the analytical XLAF model according to the ratio of the observed to predicted number of sources,
\begin{equation}
\phi_\mathrm{bin} = \phi_\mathrm{mdl} \times \dfrac{N_\mathrm{obs}}{N_\mathrm{mdl}},
\end{equation}
where $N_\mathrm{obs}$ is the number of detected sources and $N_\mathrm{mdl}$ is the number expected from  $\phi_\mathrm{mdl}$, the best-fitting XLAF model within the same bin, obtained by integrating the model over the survey sensitivity function. This method preserves the overall shape of the parametric model while providing a data-driven visualisation of the luminosity function. The uncertainties in each bin are computed assuming Poisson statistics.

\begin{figure*}[t]
    \centering
    \includegraphics[width=\linewidth]{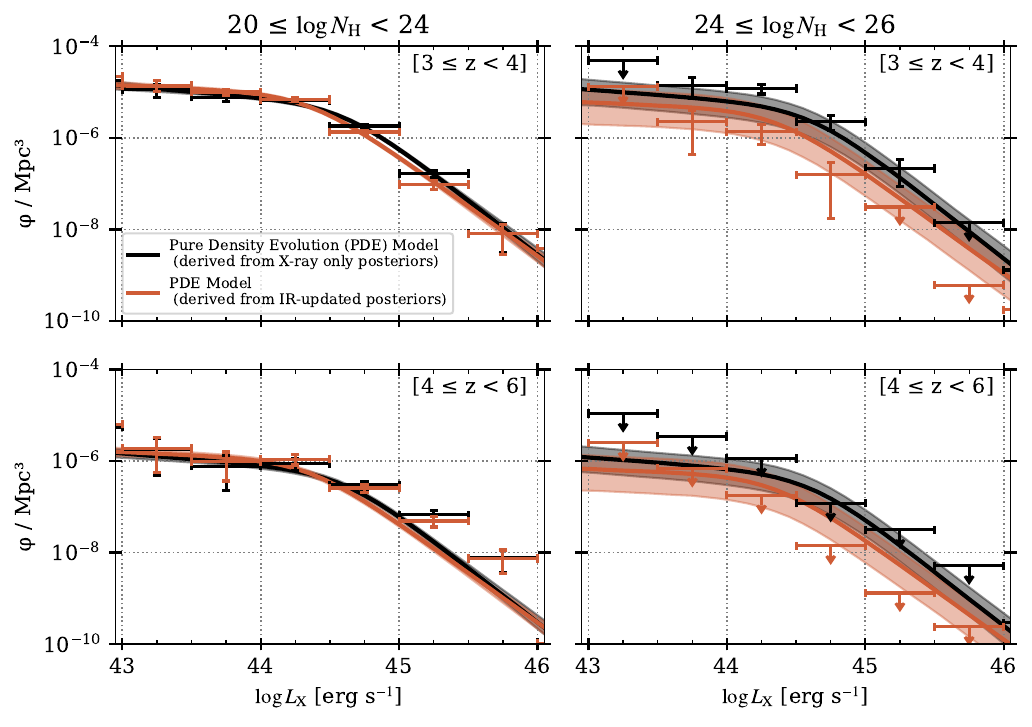}
    \caption{X-ray luminosity and absorption function for two redshift intervals ($z = 3-4$, top panels; $z = 4-6$, bottom panels) and two hydrogen column density ranges ($\log\mnh=20-24$, left panels; $\log\mnh=24-26$, right panels). Black lines and grey shaded regions indicate the results for the parametric XLAF obtained from the X-ray-only posteriors, while red lines and shaded regions correspond to those derived using the IR luminosity priors. The corresponding binned XLAFs were derived using the \citet{Miyaji2001} method.}
    \label{fig:xlaf}
\end{figure*}

Figure~\ref{fig:xlaf} presents our final XLAF estimates for two redshift intervals ($z = 3-4$ and $z = 4-6$) and two column density ranges ($\log\mnh=20-24$ and $\log\mnh=24-26$). Black solid lines and grey shaded regions show the median and $1\sigma$ uncertainty of the XLAF derived using X-ray data alone, while red lines and the corresponding shaded regions show the results obtained after including the IR-updated posteriors. In the unabsorbed and Compton-thin regimes, the results are consistent, both exhibiting similar XLF shapes, confirming the robustness of the \citet{Pouliasis2024} results. In contrast, the CTK regime shows a systematic shift towards lower space densities when the IR information is included; an expected outcome given that, as shown in Sect.~\ref{sec:lxl6}, the number of CTK sources is reduced by approximately half once the infrared constraints are applied (see also Fig.~\ref{fig:lognh_loglx_obsdist}).

The derived XLAF can also be used to estimate the intrinsic fraction of CTK AGN within our redshift range of interest. This fraction is defined as
\begin{equation}
\label{eq:fctk}
    f_\mathrm{CTK} = \dfrac{N(24\leq \log\mnh \leq 26)}{N(20\leq \log\mnh \leq 26)},
\end{equation}
where $N$ represents the number of sources within the specified comoving volume and column density range, obtained by integrating the XLAF over the relevant intervals in redshift, luminosity, and \nh. From our best-fitting XLAF model, we find an intrinsic CTK fraction of $f_\mathrm{CTK}=0.17^{+0.12}_{-0.11}$ in the redshift range $z = 3-6$. A comparison of this estimate with previous results will be discussed in Sect.~\ref{sec:discussion} below.

\section{Discussion}
\label{sec:discussion}

\subsection{Reliability of \nh\ estimates}
\label{sec:nhreliability}

A reasonable concern is whether \XMM ~and \Chandra~ that operate in a relatively soft energy band below 10 keV, can efficiently detect CTK AGN. Indeed, extending the energy range of the X-ray spectra above 10 keV e.g. using {\it NuSTAR} can provide accurate column density estimates especially at low redshifts \citep[e.g.][]{Ricci2015}. However, at the redshift range explored  in this work the energy  turnovers due to the mild CTK absorption are expected to lie at energies above 1.5 keV. These energies are where the effective area of \XMM ~and \Chandra ~is the highest. Furthermore, \textit{NuSTAR} is not sensitive enough to detect sources in the redshift range we explore in this paper. For example, \citet{Aird2015a} compiled a sample of 94 sources using the \textit{NuSTAR} extragalactic survey program \citep{Harrison2013} and 
 estimated the AGN XLF. No radio-quiet AGN have been detected above a redshift of $z = 3$. 
 Moreover, \citet{Tokayer2025} caution that \XMM~and \Chandra~spectral fits using a limited number of counts are reliable especially 
 in the case of CTK AGN. As already mentioned above, \cite{Laloux2023} 
 attempt to limit the effect of this constraint by employing mid-IR priors.
 
 In Sect.~\ref{sec:analysis} we examined the reliability of the X-ray–derived absorption estimates for our sample of CTK AGN candidates. By combining the X-ray spectral results with multiwavelength information, in particular the infrared constraints derived from SED fitting, we evaluated how robustly the X-ray data alone can identify genuine CTK sources across surveys of varying depth.
As discussed in Sect.~\ref{sec:lxl6}, the number of CTK candidates in our sample decreases from ten to three once their multiwavelength properties are examined in detail. In most cases, the derived column densities remain indicative of a heavily obscured environment, although they fall below the formal CTK threshold. This suggests that while strong absorption is present, the extreme levels implied by the initial X-ray analysis are not always robust.

The impact of introducing IR information varies across surveys. This effect is illustrated in Fig.~\ref{fig:probctk} where we show the probabilities of our selected CTK candidates of having $\log\mnh>24$ derived from the corresponding posteriors before and after updating with the IR information. Sources in the \Chandra\ Deep Fields are the least affected, owing to the significantly longer exposure times in these regions. Their spectra typically contain more than 100 net counts, allowing for a reliable detection of the Fe $\mathrm{K}_\alpha$ emission line expected in CTK AGN \citep{George1991} and providing well-constrained estimates of \nh\ based solely on the X-ray data.

In contrast, the CTK candidates identified in the shallower CCLS and XXL-N surveys have much lower-count spectra, often containing only a few tens of counts. While the overall continuum shape in these cases is consistent with high absorption, the inferred \nh\ values are considerably less secure. In such low signal-to-noise regimes, apparent CTK classifications may be driven by Poisson fluctuations near the observed energy of the Fe $\mathrm{K}_\alpha$ line rather than by genuine spectral features.

These results highlight that only high signal-to-noise X-ray observations can reliably constrain \nh\ in the CTK regime at these redshifts. For the majority of current \XMM ~ and \Chandra ~ surveys, the available depth is insufficient for a robust identification of CTK sources at high redshift based on X-ray data alone. Incorporating complementary IR information is therefore essential to obtain a reliable census of the most heavily obscured AGN.

\subsection{The CTK AGN luminosity function}
\label{sec:xlfcomparison}

Section \ref{sec:xlf} presented our estimates of the XLF for the high-redshift CTK AGN population. In that analysis, we tested the impact of incorporating IR–updated priors into the X-ray spectral fitting results. The inclusion of IR information systematically shifts the inferred space densities towards lower values, reflecting a more conservative and physically consistent view of the CTK population. The discussion that follows refers exclusively to these IR-updated results.

\begin{figure}
\centering
\resizebox{\hsize}{!}{\includegraphics{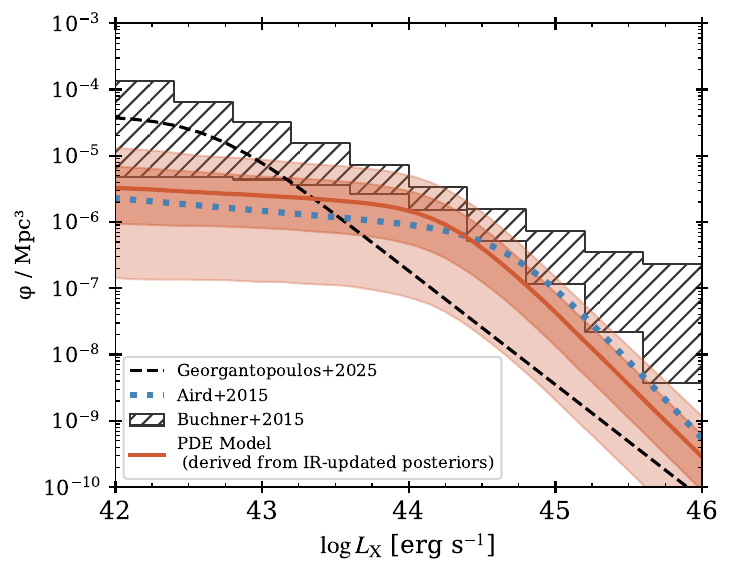}}  
\caption{X-ray luminosity function for the CTK population ($24 <\log\mnh <26$). Red symbols show our result for the parametric XLF in the 3--6 redshift range. The red solid line corresponds to the median value and shaded areas are the $1\sigma$ (darker red) and $2\sigma$ (lighter red) uncertainties. The black-dashed line is the XLF for CTK sources in the local universe ($z < 0.05$) derived by \citet{Georgantopoulos2025}. The grey, hatched area is the non-parametric XLF in the 3.1--7 redshift range estimated by \citet{Buchner2015}. The area shows the 90\% confidence region. The blue-dotted line corresponds to the \citet{Aird2015} XLF for CTK AGN in the 3--6 redshift range. 
}
\label{fig:xlf_ctk_comparison}
\end{figure}

The binned luminosity function (Fig.~\ref{fig:xlaf}) shows that our data provide meaningful constraints on the CTK population only within the $z = 3-4$ redshift interval and over the luminosity range $43.5 \leq \log\LX \leq 45$. At lower luminosities, the CTK population remains essentially unconstrained, as the current surveys lack the sensitivity required to reliably detect CTK sources below $\log\LX\sim43$ at these redshifts. Similarly, for $z = 4-6$, the XLF is constrained only by upper limits, indicating that the available data are insufficient to fully characterise the evolution of the most heavily obscured AGN at these epochs.

It should also be noted that the \citet{Pouliasis2024} sample is selected in the soft X-ray band (0.5–2~keV). At the redshifts considered here, this energy range approximately corresponds to the rest-frame hard band (2–10~keV), which may lead to the exclusion of the most extremely obscured, reflection-dominated CTK sources. Indeed, our sample does not include some known high-z CTK AGN, such as the $z = 4.762$ source in the CDF-S reported by \citet{Gilli2014}. These selection effects should be kept in mind when interpreting our results, as they imply that our parametric estimates of the X-ray luminosity and absorption function rely partly on extrapolations in luminosity and redshift regimes where direct observational constraints are limited.

Figure~\ref{fig:xlf_ctk_comparison} presents our final CTK XLF for $z = 3-6$, compared with previous studies. We include the high-redshift CTK XLFs of \citet{Buchner2015} and \citet{Aird2015}, as well as the local ($z < 0.05$) CTK XLF derived by \citet{Georgantopoulos2025}. The latter was obtained from a \swift-BAT-selected sample of local AGN, with \nh\ values measured using \textit{NuSTAR} observations. The high-z studies by \citet{Buchner2015} and \citet{Aird2015} were selected because they are based on X-ray surveys similar to ours---combining deep, pencil-beam observations with shallower, wide-area fields (using \Chandra\ and \XMM\ observations)---and employ comparable Bayesian methodologies to infer the luminosity function. The principal distinction lies in their functional forms: \citet{Buchner2015} adopted a non-parametric approach, whereas \citet{Aird2015} employed a parametric model. The latter, a flexible double power-law formulation, allows all parameters of the double power law to evolve as polynomial functions of $\log(1+z)$, thereby accommodating redshift-dependent variations in the XLF shape driven by the data. We also compared with the XLAF of \citep{Ananna2019}. This luminosity function is derived from a population synthesis model aiming to reproduce the Cosmic X-ray Background and the observed counts of CTK-AGN in surveys with \XMM, \Chandra, NuSTAR and \swift-BAT.

\begin{figure}
\centering
\resizebox{\hsize}{!}{\includegraphics{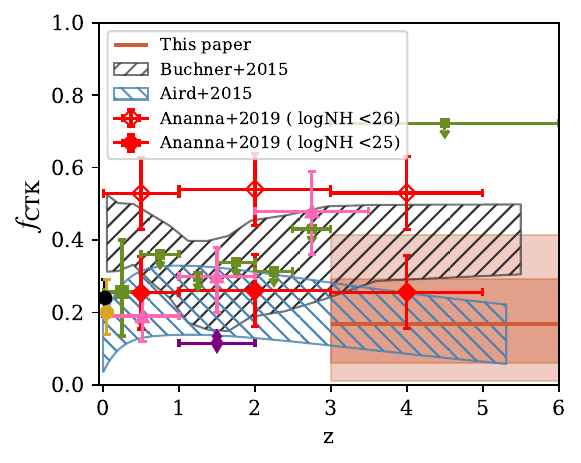}}
\caption{Evolution of the fraction of CTK sources over the total AGN population across redshift. Our results are shown with the red shaded regions that represent the $1\sigma$ (darker red) and $2\sigma$ (lighter red) confidence regions. The gray-hatched area corresponds to the \citet{Buchner2015} results (90\% confidence interval) for AGN in the $43.2 \leq \log\LX \leq 43.6$ interval. The blue-hatched area shows the \citet{Aird2015} estimates (99\% confidence interval) for sources with $\log\LX = 43.5$. The green squares show the results by \citet{Laloux2023}; quoted upper-limits are $3\sigma$. Pink triangles correspond to the results by \citet{Lanzuisi2018} and purple diamond corresponds to a study by \citet{Masini2018}. The black \citep{Georgantopoulos2025} and yellow \citep{Burlon2011} circles show the CTK fraction in the local universe using \swift-BAT selected AGN. Red diamonds show the fraction of CTK sources estimated through the integration of the \citet{Ananna2019} luminosity function. Open symbols show the result in the range ($24 \le \log\mnh < 26$), while filled symbols are for ($24 \le \log\mnh < 25$).}
\label{fig:fctk}
\end{figure}

Our results are consistent with \citet{Aird2015} and systematically lower than those reported by \citet{Buchner2015}. When compared to the local Universe, our XLF indicates a substantially lower space density of CTK sources at low luminosities ($\log\LX < 43$), while all studies show an increasing density towards higher luminosities ($\log\LX > 44$). Interestingly, the local XLF of \citet{Georgantopoulos2025} aligns closely with the low-luminosity end of the \citet{Buchner2015} results. 

The discrepancy between our results and those of \citet{Buchner2015} may stem from methodological differences. Their analysis relied solely on X-ray data, without the benefit of IR constraints, which, as we have shown, tend to reduce the inferred CTK number densities. However, the consistency between our IR-informed results and those of \citet{Aird2015}, which were also based exclusively on X-ray data, suggests that additional factors may be at play. A key distinction is that \citet{Buchner2015} derived individual source properties through full spectral modelling, similar to our approach, whereas \citet{Aird2015} employed broad-band count-rate analyses. It is likely that the latter method is less prone to spurious identification of CTK sources, as it does not rely on detailed spectral features, such as the Fe $\mathrm{K}_\alpha$ line, that can be affected by low-count statistical fluctuations.

Overall, our results suggest that the space density of CTK AGN at $z = 3-6$ is lower than previously inferred from X-ray data alone, although the uncertainties remain substantial due to limited statistics and selection biases. The consistency between our IR-informed analysis and independent parametric estimates reinforces the robustness of our approach, while also highlighting the need for deeper, high-energy observations--such as those anticipated from \textit{NewAthena} and other future missions--to fully characterise the obscured growth of supermassive black holes in the early Universe.

\subsection{The evolution of the CTK fraction}
\label{sec:fctkcomparison}

Figure~\ref{fig:fctk} compares our estimates of the CTK fraction\footnote{Our results for the CTK fraction are derived from our estimate of the XLAF (see Eq.~\ref{eq:fctk}), which takes into account the likelihood of sources being detected/undetected as functions of, $\LX, z$, and \nh\ (see Appendix~\ref{app:xlaf_details}.}
 in the redshift range $z = 3-6$ with previous measurements spanning from the local Universe up to $z \sim 6$. The results from \citet{Buchner2015} and \citet{Aird2015} are derived from the luminosity functions presented in Sect.~\ref{sec:xlfcomparison}. The estimates of \citet{Lanzuisi2018} are based on a study of CTK AGN selected in the CCLS, while those of \citet{Laloux2023} also rely on CCLS data but employ a non-parametric approach to model the X-ray luminosity and absorption function of the AGN population.
In the latter work, X-ray spectral fitting was performed for individual sources using IR–derived priors. The measurement by \citet{Masini2018} originates from a \textit{NuSTAR} survey of the UKIDSS Ultra Deep Survey field and represents the observed CTK fraction, which can be considered a lower limit to the intrinsic value. For comparison, we also include two estimates of the CTK fraction for the local Universe from \citet{Burlon2011} and \citet{Georgantopoulos2025}.  Additionally, we have incorporated the CTK fractions estimated through the integration of the luminosity function estimated by \citet{Ananna2019}. We calculated the CTK fraction across two different \nh\ intervals ($24 \leq \log\mnh \leq 25$, and $24 \leq \log\mnh \leq 26$).

Among these studies, the results of \citet{Buchner2015} and \citet{Lanzuisi2018} stand out for their remarkably high CTK fraction of approximately 40-50\%, among the largest values reported in the literature at high redshift. Such high fractions are, however, in clear tension with both our findings and those of other works. For example, \citet{Aird2015} reported lower CTK fractions that are broadly consistent with our results in the $z = 3-6$ range, although they noted that their highest-redshift bins may be affected by limited statistics. The results of \citet{Ananna2019} show no evolution with redshift and, for the population with $\log\mnh <25$, the CTK fraction is about 30 per cent, consistent with previous results in similar redshift ranges. The estimate for the high redshift range ($3<z<5$) aligns with our estimate of the CTK fraction at high redshifts. However, when considering the population with $\log\mnh > 25$, the CTK fraction from \citet{Ananna2019} reaches $\sim50\%$ that is similar to the highest values reported in the literature\citep{Lanzuisi2018}.  This discrepancy may arise from the limited sensitivity to this highly absorbed population in the X-ray energies examined by these studies.

At intermediate redshifts ($0.5 < z < 2.5$), most studies converge on a CTK fraction of 15–30\%, consistent with estimates for the local Universe. Overall, the emerging picture suggests no compelling evidence for strong evolution of the CTK fraction up to $z\approx 5-6$. The only study reporting a significant increase is that of \citet{Lanzuisi2018}, who found a rise from $\sim20\%$ at low redshift to nearly 50\% at $z\sim3$. These results appear to be in tension with the more recent analysis by \citet{Laloux2023} of the same field, whose upper limits around $z\sim3$ favour a lower CTK fraction than that inferred by \citet{Lanzuisi2018}.

In contrast, the evolution of the total obscured AGN population (including Compton-thin sources) is well established \citep[e.g.][]{Vito2018,Signorini2023,Peca2023,Pouliasis2024}, showing a pronounced increase towards higher redshifts. This trend may be explained by the findings of \citet{Gilli2022}, who proposed that the rise in the obscured AGN population is primarily linked to the interstellar medium (ISM) of the host galaxies. Their model suggests that ISM-driven obscuration rarely reaches the CTK regime at these redshifts. Instead, the ISM is expected to dominate obscuration within the Compton-thin range, while the relative contribution of CTK sources remains approximately constant across cosmic time.

These findings support a scenario in which the most heavily obscured, CTK phase of AGN activity represents a relatively stable component of black hole growth. The strong redshift evolution observed in the overall obscured population is therefore likely driven by increasing Compton-thin obscuration associated with the evolving ISM of high-redshift galaxies \citep{Scoville2014,Scoville2017}.

\section{Conclusions}
\label{sec:conclusions}

We investigated the population of CTK AGN at high redshift ($z \geq 3$) using the \citet{Pouliasis2024} sample, one of the largest X-ray–selected AGN datasets currently available in this regime. From this sample, we identified a subsample of ten high-probability CTK candidates based on the Bayesian X-ray spectral fits of \citet{Pouliasis2024} and examined their multiwavelength properties through SED analysis to assess the robustness of their CTK classifications. While the SEDs are generally consistent with heavily obscured AGN, a substantial fraction of the sources exhibit X-ray luminosities significantly higher than expected from the established correlations between $\LX$ and $\LS$. Most of these outliers have low-count X-ray spectra, suggesting that the discrepancy arises from an overestimate of \nh.

To address this, we applied the methodology of \citet{Laloux2023} to update the X-ray posterior distributions with the IR constraints derived from our SED analysis, yielding more physically consistent X-ray properties. After this update, only three sources retained a high probability of being CTK, while the others still showed substantial absorption but below the nominal CTK threshold.

We then derived the X-ray luminosity function (XLF) for CTK AGN in parametric form, using the complete sample and methodology of \citet{Pouliasis2024}. The individual X-ray posteriors were updated using $\LS$ upper limits estimated from \textit{Spitzer}-MIPS $24\,\microns$ photometry to obtain more robust values for $\LX$ and \nh. Incorporating IR constraints systematically reduces the inferred space density of CTK AGN, resulting in a more conservative and physically consistent description of the population, while leaving the unabsorbed and Compton-thin XLF largely unaffected. The resulting CTK XLF provides meaningful constraints for $z = 3-4$  and $43.5 \lesssim \log\LX \lesssim 45$, whereas at lower luminosities and higher redshifts the data remain limited by current survey sensitivities.

Our derived CTK XLF is consistent with that of \citet{Aird2015} and lower than the estimates of \citet{Buchner2015}, underscoring the importance of multiwavelength constraints in mitigating degeneracies in X-ray spectral fitting. We find an intrinsic CTK fraction of $f_\mathrm{CTK}=0.17^{+0.12}_{-0.11}$ in the redshift range $z = 3-6$. Comparisons with local and intermediate-redshift measurements indicate that the CTK space density remains broadly stable with redshift, in contrast to the overall obscured AGN population (including Compton-thin sources), which shows strong evolution towards higher redshift. This trend likely reflects increasing interstellar medium obscuration in high-redshift galaxies, while the intrinsic CTK fraction remains approximately constant across cosmic time.

Overall, our results support a scenario in which the CTK phase represents a persistent but relatively rare mode of black hole growth, contributing a stable fraction to the accretion history of the Universe. Future high-sensitivity, broad-band X-ray observatories such as \textit{NewAthena} and \textit{HEX-P} will be essential to further constrain this population, extending current analyses to lower luminosities and providing a more robust census of the most heavily obscured AGN at early cosmic epochs.

\begin{acknowledgements}
The research leading to these results has received funding from the Hellenic Foundation for Research and Innovation (HFRI) project ``4MOVE-U'' grant agreement 2688, which is part of the programme ``2nd Call for HFRI Research Projects to support Faculty Members and Researchers''.
EP and IG acknowledge ACME, a project funded by the European Union's Horizon Europe Research and Innovation programme under Grant Agreement 101131928.
Based on observations obtained with \XMM, an ESA science mission with instruments and contributions directly funded by ESA member states and NASA.
This research has made use of data obtained from the \Chandra\ Data Archive and the \Chandra\ Source Catalogue, and software provided by the \Chandra\ X-ray Center (CXC) in the application packages CIAO and Sherpa.
This research has made use of the NASA/IPAC Infrared Science Archive, which is funded by the National Aeronautics and Space Administration and operated by the California Institute of Technology.
This work is based on observations made with the \textit{Spitzer} Space Telescope, which is operated by the Jet Propulsion Laboratory, California Institute of Technology under a contract with NASA. 
This work made use of Astropy:\footnote{\url{http://www.astropy.org}} a community-developed core Python package and an ecosystem of tools and resources for astronomy \citep{astropy:2013, astropy:2018, astropy:2022}.
The plots in this publication were produced using Matplotlib, a Python library for publication-quality graphics \citep{matplotlib}.

\end{acknowledgements}

\bibliographystyle{aa}
\bibliography{references} 

\begin{appendix}

\onecolumn

\section{Calculation of the X-ray luminosity and absorption function}
\label{app:xlaf_details}

This appendix describes the parametric modelling of the X-ray luminosity and absorption functions used to characterise the high-redshift AGN population analysed in this work. We outline the adopted analytical forms for the luminosity and absorption distributions, their assumed redshift evolution, and the Bayesian framework employed to constrain the model parameters.

\subsection*{Parametric form for the XLAF}

We modelled the differential X-ray luminosity function assuming a broken power-law form, which has been shown to describe well the shape of the local AGN population. It is defined as:
\begin{equation}
\phi(\LX,z=0) = \dv{\Phi(\LX,z=0)}{\log\LX} =
A \times \left[\left(\dfrac{\LX}{L_*}\right)^{\gamma_1} + \left(\dfrac{\LX}{L_*}\right)^{\gamma_2}\right]^{-1},
\end{equation}
where $A$ is the normalisation factor, $L_*$ is the characteristic luminosity break, and $\gamma_1$ and $\gamma_2$ are the slopes of the power law before and after $L_*$, respectively \citep{Miyaji2000,Hasinger2005}.

To account for cosmic evolution, we adopted a pure density evolution model \citep[PDE;][]{Schmidt1968}, in which the luminosity function varies with redshift according to:
\begin{equation}
\phi(\LX,z) = \dv{\Phi(\LX,z=0)}{\log\LX} \times e(z),
\end{equation}
where $e(z)$ characterises the redshift evolution of the local luminosity function and is expressed as
\begin{equation}
e(z) = \qty(\dfrac{1+z}{1+z_c})^\pden.
\end{equation}

Given our definition for the XLAF (see Eq.~\ref{eq:xlaf}, Sect.~\ref{sec:xlf}), we need to assume a functional form for the absorption function, $\fabs$.Following the methodology of \citet{Ueda2003}, we modelled the absorption function using piecewise constant functions across discrete \nh\ bins. Given the redshift range of our study and the energy coverage of \XMM\ and \Chandra, X-ray absorption cannot be reliably constraint for column densities below $\sim 10^{23}~\mathrm{cm^{-2}}$. We therefore defined the absorption function in three \nh\ intervals as follows:
\begin{equation}
\fabs(\mnh, z, \LX) = 
\begin{cases}
\dfrac{1}{3} - \dfrac{\varepsilon}{3(1 + \varepsilon)} \psi(z,\LX) & [20 \leq \log \mnh < 23] \\
\dfrac{\varepsilon}{1 + \varepsilon} \psi(z,\LX) & [23 \leq \log \mnh < 24]\\
\dfrac{\fctk}{2}~\psi(z,\LX) & [24 \leq \log \mnh < 26]
\end{cases}
\end{equation}
In this parametrisation, $\varepsilon$ represents the ratio between the number of sources with $23 \leq \log\mnh < 24$ and those with $22 \leq \log\mnh < 23$, while $\fctk$ denotes the relative fraction of CTK sources with respect to absorbed Compton-thin AGN \citep{Vijarnwannaluk2022}.

The quantity $\psi(z, \LX)$ corresponds to the fraction of absorbed Compton-thin AGN relative to the total AGN population. It incorporates both the redshift and luminosity dependence and is parametrised as a linear function of $\log\LX$:
\begin{equation}
\label{beta}
\psi(z,\LX)=\min(\psi_{\rm max}, \max(\psi_{43.75}(z) - C(\log\LX - 43.75), \psi_{\rm min})),
\end{equation}
where we adopt $\psi_{\rm min} = 0.2$ and $\psi_{\rm max} = 0.99$. The parameter $C$ controls the luminosity dependence, while $\psi_{43.75}(z)$ represents the absorption fraction for AGN with $\log\LX = 43.75$ at a given redshift.

For $z < 2$, $\psi_{43.75}(z)$ is well constrained \citep{Ueda2014}, whereas above this redshift it is typically assumed constant \citep[e.g. $2\leq z <3$;][]{Vijarnwannaluk2022}. In this work, we extended the definition of $\psi_{43.75}(z)$ for $z \geq 3$ as:
\begin{equation}
\psi_{43.75}(z \geq 3) = \psi_3 \times (1+z)^{a_2},
\end{equation}
where $\psi_3$ denotes the absorption fraction at $\log\LX = 43.75$ and $z = 3$, and $a_2$ is the redshift evolution index.

\begin{figure*}
    \centering
    \includegraphics[width=\linewidth]{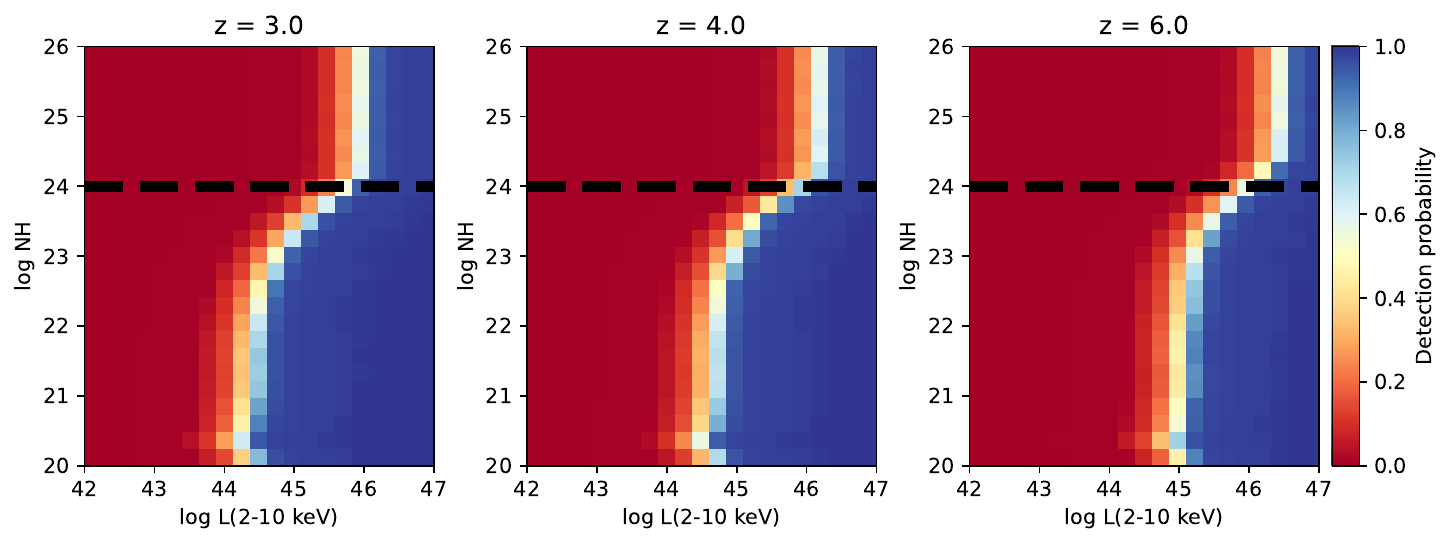}
    
    \caption{Sensitivity maps ($\Omega$) for our survey. Each panel show the detection probability in the $\log\LX, \log\mnh$ plane at three redshift values (3.0, 4.0, 6.0).}
    \label{fig:omega_maps}
\end{figure*}

\subsection*{Bayesian inference framework}

We employed a Bayesian approach to estimate simultaneously the parametric form of the XLF and the absorption function. Given a dataset of $n$ observations, $D = \{d_i; i = 1, \ldots, n\}$, and a model for the XLF described by a set of parameters $\vb*{\Theta}$, Bayes’ theorem gives:
\begin{equation}
P(\vb*{\Theta} | D) = \dfrac{P(D|\vb*{\Theta}) P(\vb*{\Theta})}{P(D)},
\end{equation}
where $P(\vb*{\Theta} | D)$ is the posterior probability of the model given the data,  
$\La = P(D | \vb*{\Theta})$ is the likelihood of obtaining the data for a given model, $P(\vb*{\Theta})$ is the prior probability of the model parameters,  
and $P(D) = \int P(\vb*{\Theta} | D) \dd\vb*{\Theta}$ is the evidence.

\begin{table}
\caption{Prior limits and best-fitting values for the free parameters of the X-ray luminosity and absorption functions. Results are shown for both cases: using X-ray–only posteriors for the individual sources, and using posteriors updated with infrared (IR) information.}
\label{tab:xlaf_params}
\centering
\begin{tabular}{clccc}
\hline\hline
   & Parameter  & Prior limits & \multicolumn{2}{c}{Best-fit value} \\
\cmidrule(lr){4-5}   
   &            &              &   X-ray                 &   X-ray and IR   \\
\hline
\multirow{5}{50pt}{Luminosity function}  
   & $\log A$      & [-5, -3]  & $-4.25^{+0.10}_{-0.10}$ & $-4.03^{+0.09}_{-0.10}$ \\
   & $\log L_*$    & [42,  46] & $44.60^{+0.08}_{-0.07}$ & $44.36^{+0.08}_{-0.07}$ \\
   & $\gamma_1$    & [-2,  2]  & $0.25^{+0.09}_{-0.09}$  & $0.13^{+0.10}_{-0.10}$  \\
   & $\gamma_2$    & [1,  6]   & $2.35^{+0.17}_{-0.16}$  & $2.16^{+0.13}_{-0.12}$  \\
   & $\pden$       & [-10, -3] & $-8.53^{+0.49}_{-0.47}$ & $-8.32^{+0.47}_{-0.46}$ \\
\hline
\multirow{5}{50pt}{Absorption function}  
   & $\varepsilon$ & [0, 10]   & $2.841^{+0.37}_{-0.31}$ & $2.14^{+0.29}_{-0.27}$  \\
   & $\fctk$       & [0, 10]   & $0.35^{+0.28}_{-0.25}$  & $<0.32$                 \\
   & $\psi_3$      & [0, 1]    & $0.60^{+0.39}_{-0.25}$  & $0.99^{+0.29}_{-0.98}$  \\
   & $C$           & [-10, 10] & $-3.0^{+6.0}_{-6.8}$    & $-2.4^{+8.9}_{-4.2}$    \\
   & $a_2$         & [0, 15]   & $10.6^{+3.6}_{-5.3}$    & $12.1^{+2.7}_{-6.0}$    \\
\hline
\end{tabular}
\tablefoot{The normalisation, $A$, and the break luminosity, $L_*$, are given in units of $\mathrm{Mpc^{-3}}$ and $\mathrm{erg\,s^{-1}}$, respectively. The quoted best-fitting values and uncertainties correspond to the mode of the posterior distribution and the $1\sigma$ credible intervals around the mode.}
\end{table}

To sample the posterior distribution of the model parameters, we used the nested-sampling Monte Carlo algorithm MLFriends \citep{Buchner2016,Buchner2017}, implemented in the \texttt{UltraNest} package. Nested sampling enables simultaneous exploration of the posterior distribution and computation of the Bayesian evidence, which allows a direct comparison between alternative XLF models through Bayes factors. This approach also provides a rigorous treatment of the uncertainties in the X-ray spectral parameters and photometric redshifts of the individual sources included in the analysis.

\begin{figure*}
    \centering
    \includegraphics[width=\linewidth]{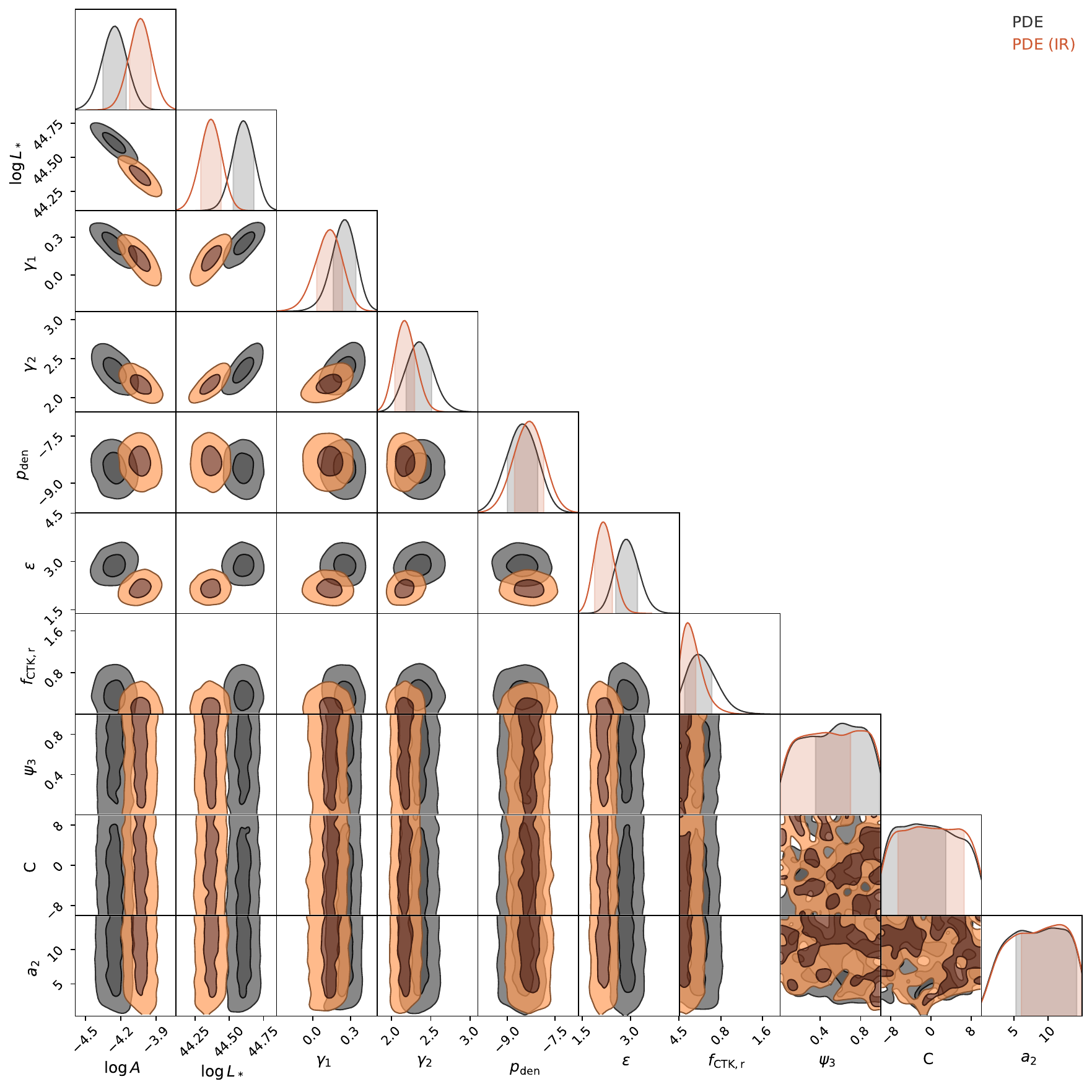}
    \caption{1D and 2D marginal posterior distributions for the XLAF parameters. Black: X-ray only posteriors; red: posteriors updated with IR data.}
    \label{fig:xlaf_params}
\end{figure*}

We adopted flat (uniform or log-uniform) priors for all free parameters, spanning ranges broad enough to encompass the values reported in previous studies. Table~\ref{tab:xlaf_params} lists the adopted prior limits for each parameter of the XLAF model.

Following the formulation of \citet{Loredo2004}, the likelihood of observing a given dataset can be expressed as the product of the probabilities of detecting each individual source multiplied by the probability of not detecting any additional sources. The corresponding log-likelihood, following \citealt{Buchner2015}, is:
\begin{equation}
\label{eq:xlflikelihood}
\ln \La = -\lambda + \sum_i \ln\iiint P_i(\LX, z, \mnh | \vb*{\Theta})~\dv{V}{z}\dlog\mnh~\dlog\LX~\dd z,
\end{equation}
where $\lambda$ represents the expected number of detected sources in a Poisson process for an XLF model with parameters $\vb*{\Theta}$:
\begin{equation}
\label{eq:expectedsources}
\lambda = \iiint \phi_\mathrm{abs}(\LX, z, \mnh | \vb*{\Theta}) \Omega(\LX, z, \mnh)~\dv{V}{z}\dlog\mnh~\dlog\LX~\dd z,
\end{equation}
and $\phi_{\mathrm{abs}} = \phi \times \fabs$, where $\phi$ and $\fabs$ are the luminosity and absorption functions defined above. $\Omega(\LX, z, \mnh)$ is the survey sensitivity function, for which we adopted the values calculated by \citet{Pouliasis2024}.

The term $P_i$ in Eq.~\ref{eq:xlflikelihood} is defined as:
\begin{equation}
P_i(\LX, z, \mnh | \vb*{\Theta}) = p(d_i | \LX, z, \mnh)~\phi_\mathrm{abs}(\LX, z, \mnh | \vb*{\Theta})~\Omega(\LX, z, \mnh).
\end{equation}
where $p(d_i|\LX, z, \mnh)$ is the probability that source $i$ has X-ray properties $(L_{\mathrm{X}}, z, N_{\mathrm{H}})$, given by the posterior distributions obtained from the Bayesian X-ray spectral fitting. Alternatively, the IR-updated posteriors discussed in Sect.~\ref{sec:lxl6} can be used here. The inclusion of $\Omega$ within this term accounts for the loss of information caused by differences between X-ray source detection and spectral fitting procedures \citep[see][Appendix~A, for a detailed discussion]{Buchner2015}.

The integral in Eq.~\ref{eq:xlflikelihood} is evaluated using importance-sampling integration techniques \citep{Kloek1978,Press2007}. The integration limits adopted for $z$, $\log\LX$, and $\log\mnh$ are [3, 6], [42, 47], and [20, 26], respectively.

Our combined parametrisation of the luminosity and absorption functions includes ten free parameters. Table~\ref{tab:xlaf_params} lists the best-fitting parameter values and their uncertainties for the assumed PDE model. Figure~\ref{fig:xlaf_params} shows the one-dimensional (diagonal panels) and two-dimensional marginal posterior distributions of the XLAF parameters. Results based on X-ray–only posteriors are shown in grey, while those incorporating IR-updated posteriors are shown in red. Most luminosity-function parameters remain consistent within $2\sigma$ between the two cases, whereas the inclusion of IR data shifts the absorption-related parameters $\varepsilon$ and $\fctk$ toward lower values. The parameters governing the redshift evolution of the absorption function ($\psi_3$, $C$, and $a_2$) remain poorly constrained, indicating that our dataset does not provide significant evidence for absorption-function evolution within the studied redshift range.

\section{X-ray properties}
\label{xrays}

\begin{figure*}[b]
\centering
\begin{subfigure}[t]{0.49\textwidth}
    \centering
    \includegraphics[width=\linewidth, trim={0 0 0 1.5cm},clip]{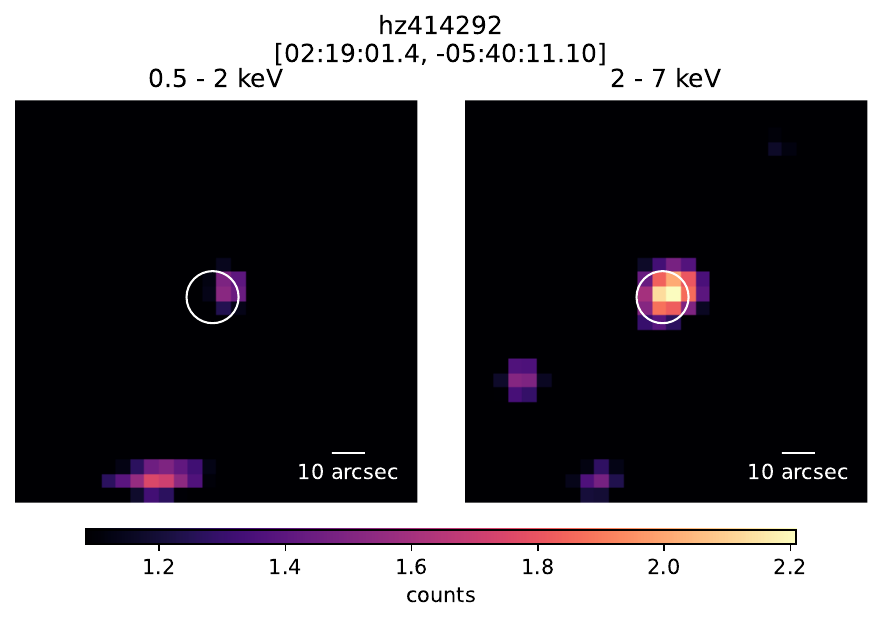}
    \includegraphics[width=\linewidth]{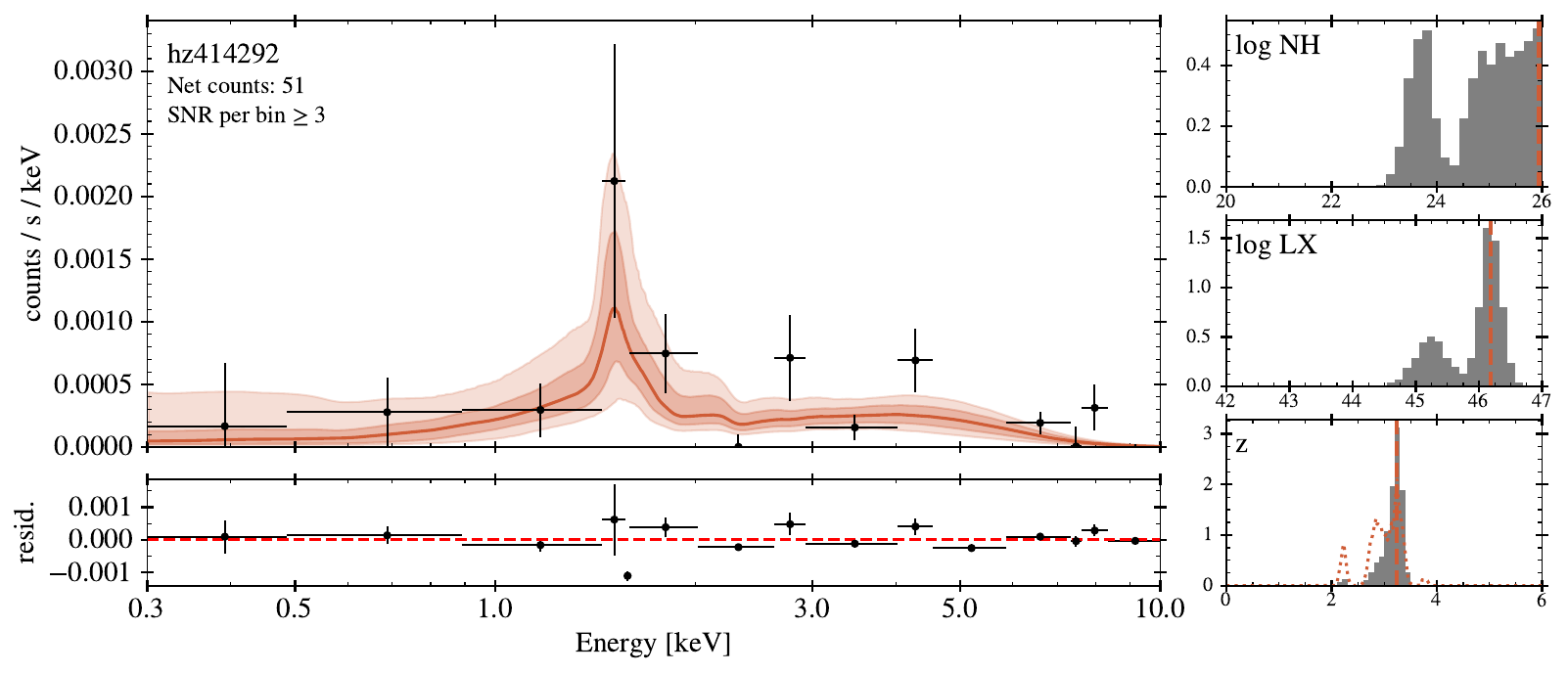}
    \label{fig:xrdata_hz414292}
\end{subfigure}
~
\begin{subfigure}[t]{0.49\textwidth}
    \centering
    \includegraphics[width=\linewidth, trim={0 0 0 1.5cm},clip]{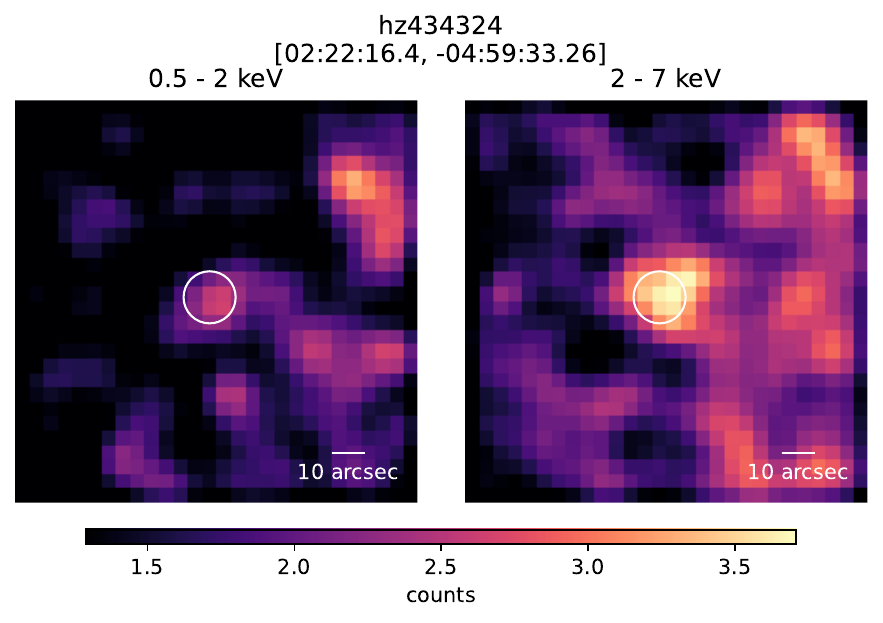}
    \includegraphics[width=\linewidth]{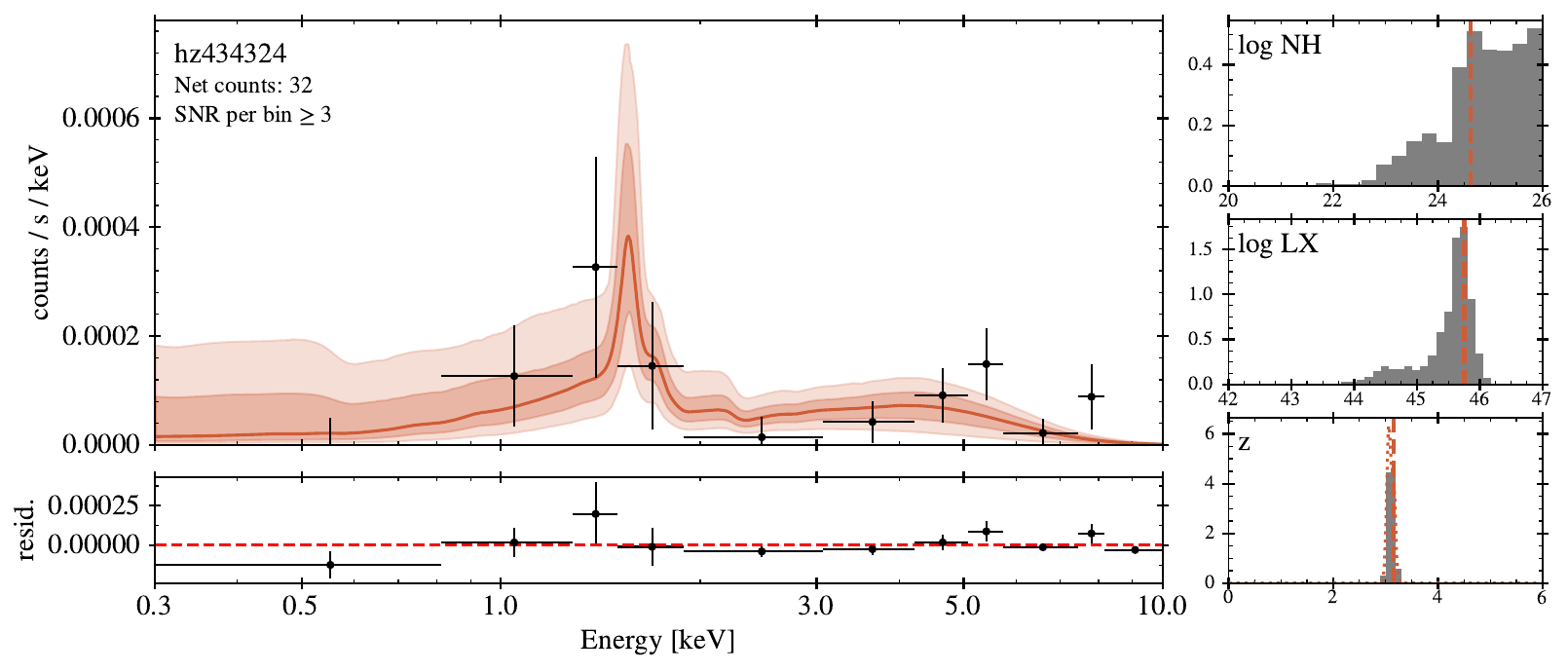}
    \label{fig:xrdata_hz434324}
\end{subfigure}

\caption{\XMM\ cutouts and spectra for the selected CTK candidates. \textbf{Top:} Two arc-minutes \XMM-EPIC cutouts in the 0.5-2~keV (left) and 2-7~keV (right) bands. \textbf{Bottom:} Co-added, background-subtracted \XMM-EPIC spectrum. To improve visualization, the spectrum is binned. The red-shaded areas show the one- and two-sigma uncertainties for the best-fit source model (red, solid line). On the right column we plot the posterior distribution for $\log \mnh$ (top), $\log \LX$ (middle), and redshift (bottom). The red, dashed lines show the mode of each distribution.}
\label{fig:xrdata_xmm}
\end{figure*}

This appendix presents, in Figs.~\ref{fig:xrdata_xmm} and \ref{fig:xrdata_chandra}, the X-ray cutouts and spectra for the ten CTK AGN candidates included in our sample. For each source, we show the \Chandra\ or \XMM\ images in the observed 0.5–2 and 2–7~keV bands, together with the corresponding X-ray spectrum and best-fitting model derived from the Bayesian spectral analysis. We also display the marginalised posterior distributions for three key parameters: $\mnh, \LX$ and $z$. For sources with spectroscopic redshifts, the z distribution is represented by a single vertical red line.

These figures illustrate the data quality and spectral characteristics of the selected CTK candidates, highlighting the range of absorption levels and luminosities found within the sample. They also provide a visual summary of the Bayesian modelling results, allowing a direct assessment of the uncertainties and degeneracies that affect the determination of \nh\ and $\LX$ for individual sources.

\begin{figure*}
\centering
\begin{subfigure}[t]{0.49\textwidth}
    \centering
    \includegraphics[width=\linewidth, trim={0 0 0 1.5cm},clip]{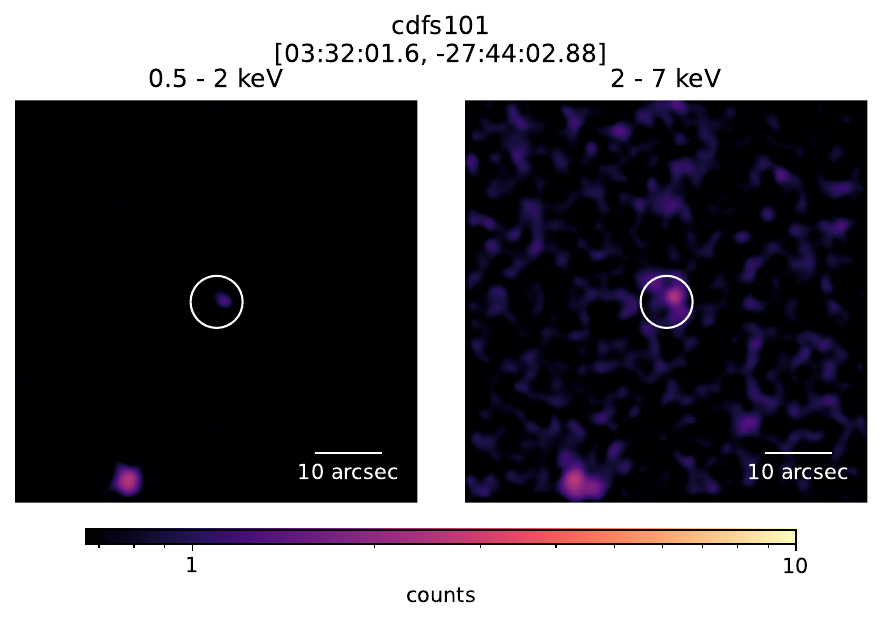}
    \includegraphics[width=\linewidth]{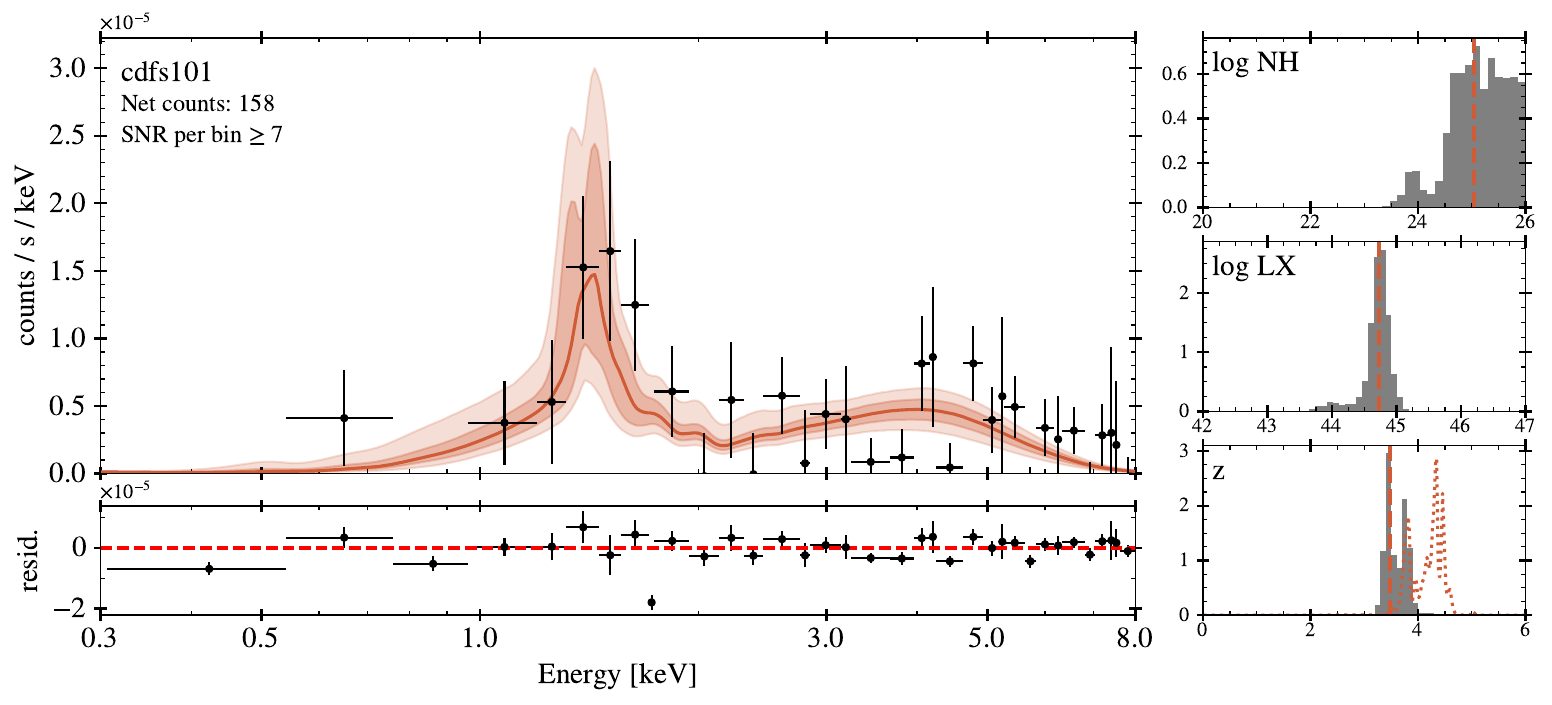}
    \label{fig:xrdata_cdfs101}
\end{subfigure}
~
\begin{subfigure}[t]{0.49\textwidth}
    \centering
    \includegraphics[width=\linewidth, trim={0 0 0 1.5cm},clip]{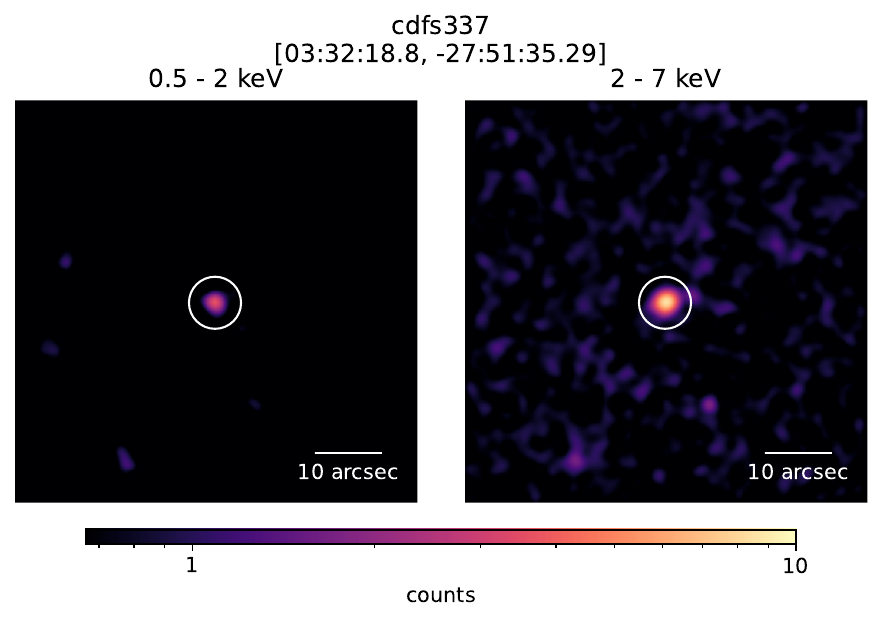}
    \includegraphics[width=\linewidth]{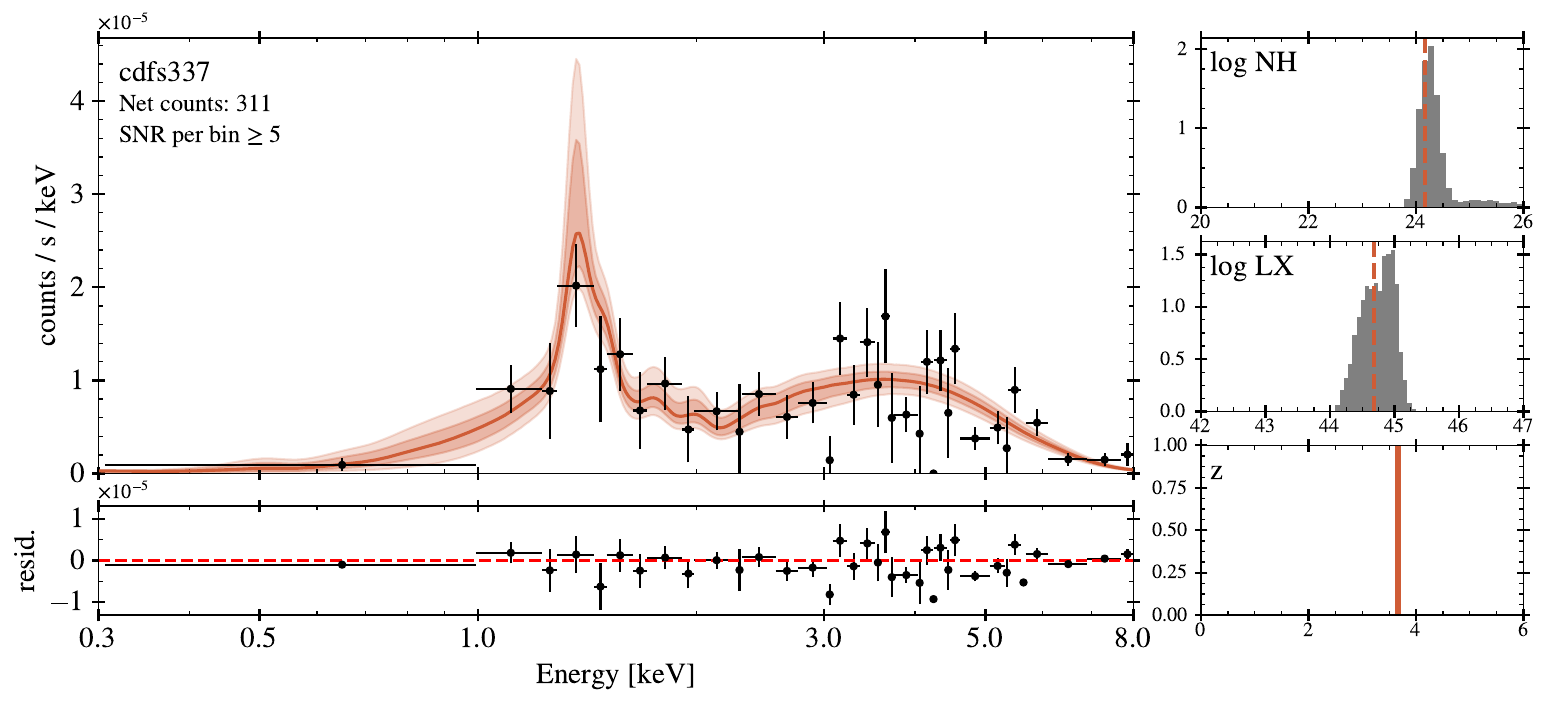}
    \label{fig:xrdata_cdfs337}
\end{subfigure}

\begin{subfigure}[t]{0.49\textwidth}
    \centering
    \includegraphics[width=\linewidth, trim={0 0 0 1.5cm},clip]{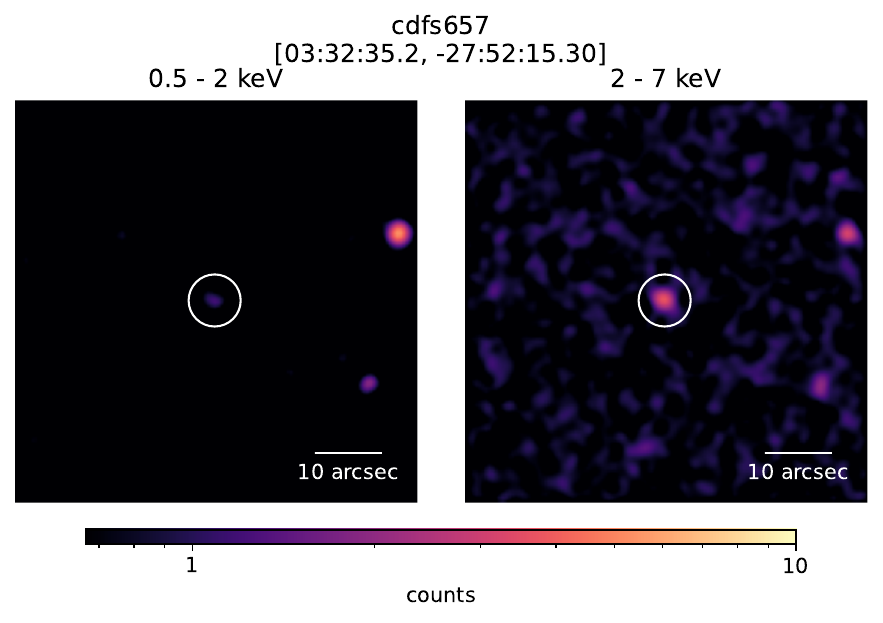}
    \includegraphics[width=\linewidth]{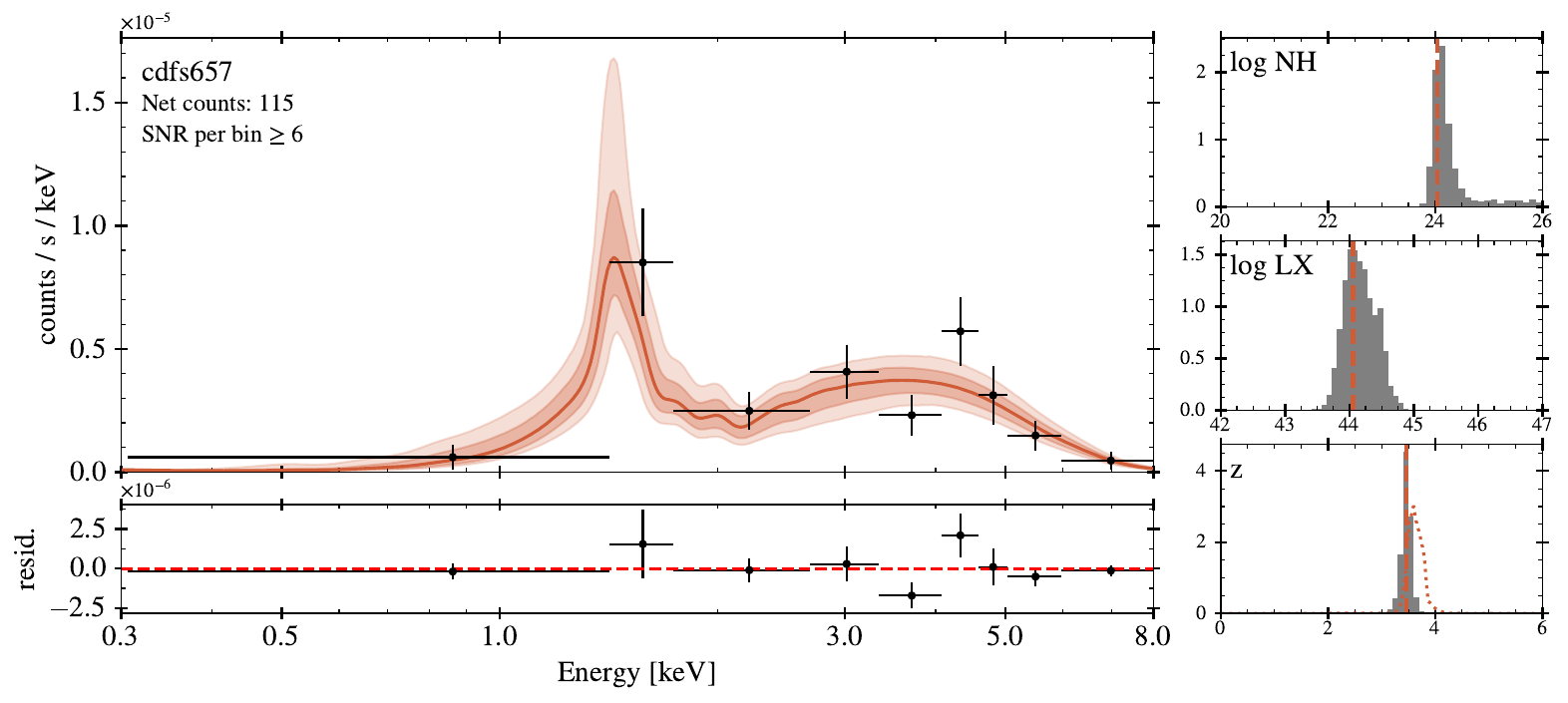}
    \label{fig:xrdata_cdfs657}
\end{subfigure}
~
\begin{subfigure}[t]{0.49\textwidth}
    \centering
    \includegraphics[width=\linewidth, trim={0 0 0 1.5cm},clip]{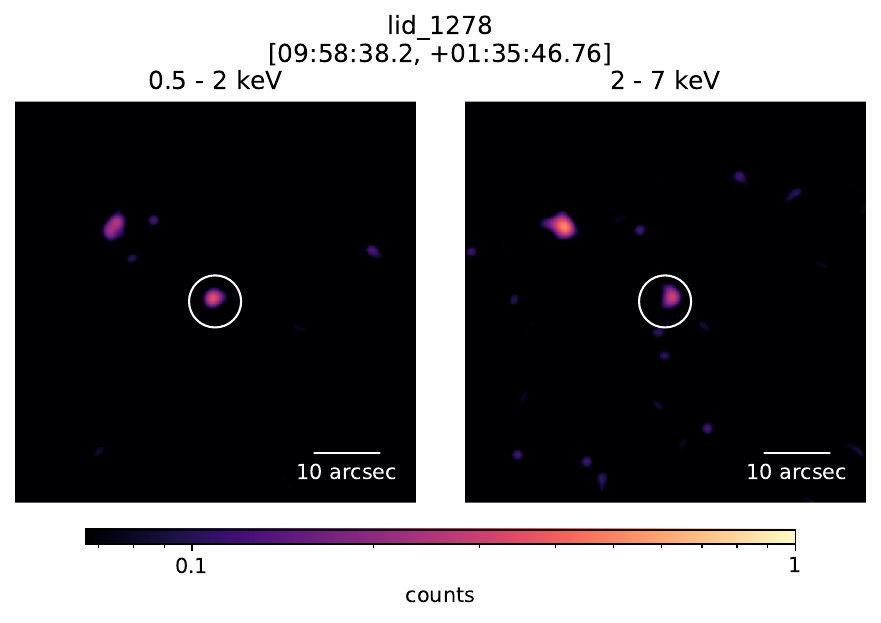}
    \includegraphics[width=\linewidth]{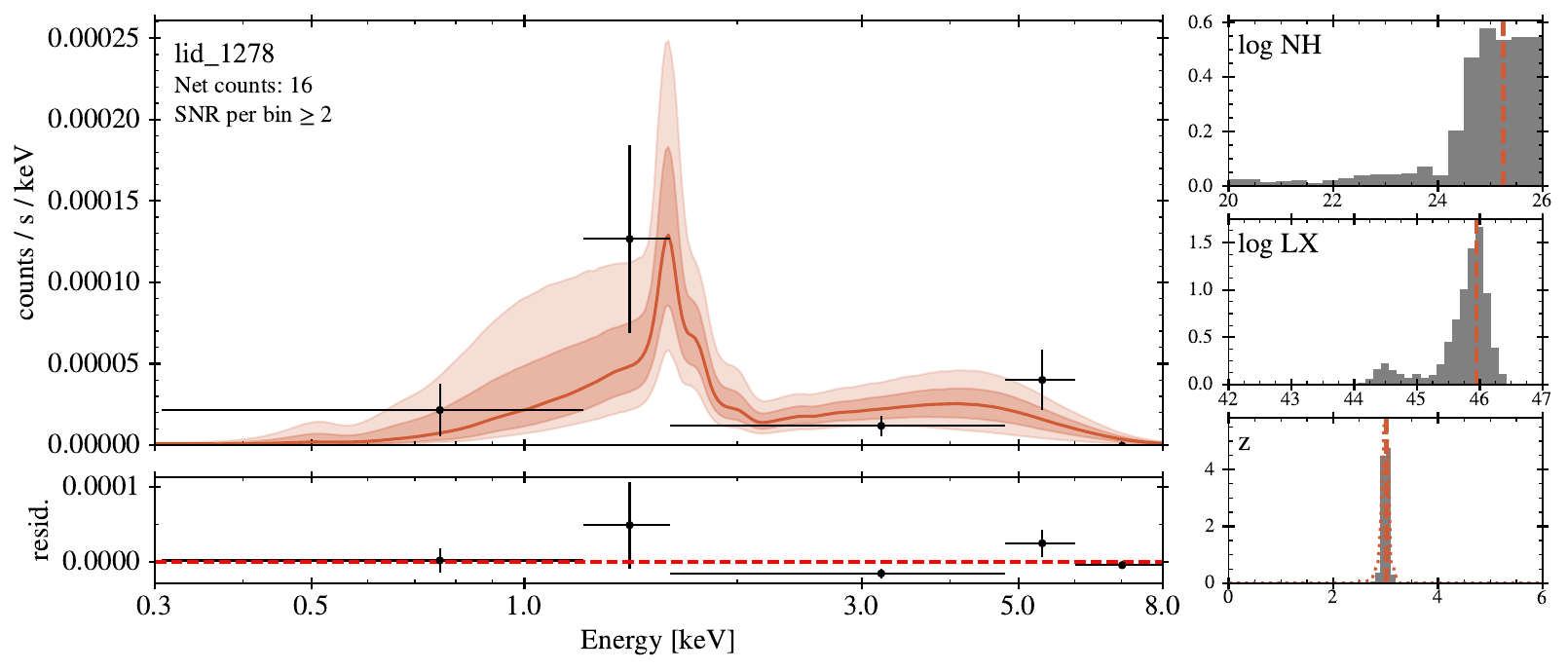}
    \label{fig:xrdata_lid1278}
\end{subfigure}

\caption{\Chandra\ cutouts and spectra for the selected CTK candidates. \textbf{Top:} One arc-minutes \Chandra-ACIS cutouts in the 0.5-2~keV (left) and 2-7~keV (right) bands. \textbf{Bottom:} Co-added, background-subtracted \Chandra-ACIS spectrum. To improve visualization, the spectrum is binned. The red-shaded areas show the one- and two-sigma uncertainties for the best-fit source model (red, solid line). On the right column we plot the posterior distribution for $\log \mnh$ (top), $\log \LX$ (middle), and redshift (bottom). The red, dashed lines show the mode of each distribution.}
\label{fig:xrdata_chandra}
\end{figure*}

\begin{figure*}
\ContinuedFloat
\centering
\begin{subfigure}[t]{0.49\textwidth}
    \centering
    \includegraphics[width=\linewidth, trim={0 0 0 1.5cm},clip]{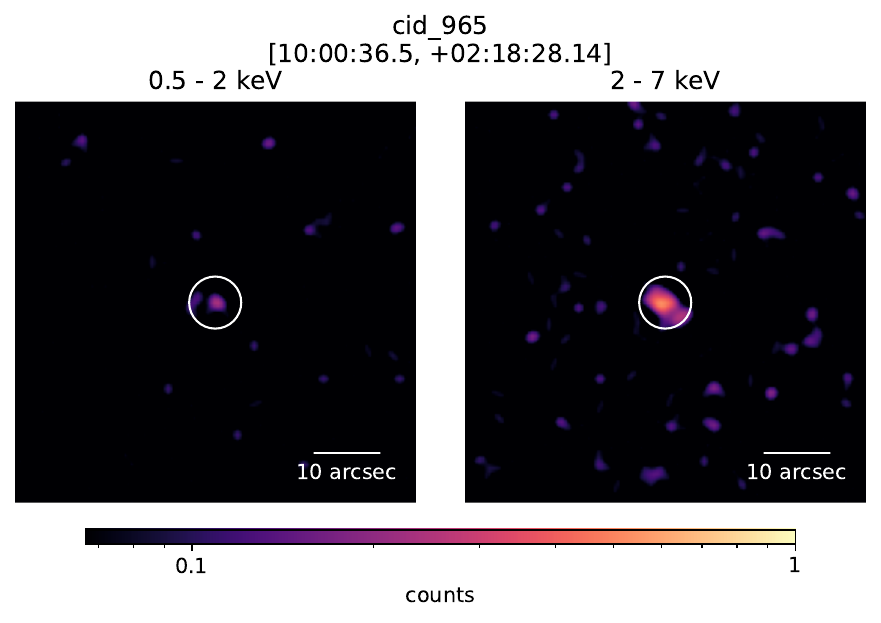}
    \includegraphics[width=\linewidth]{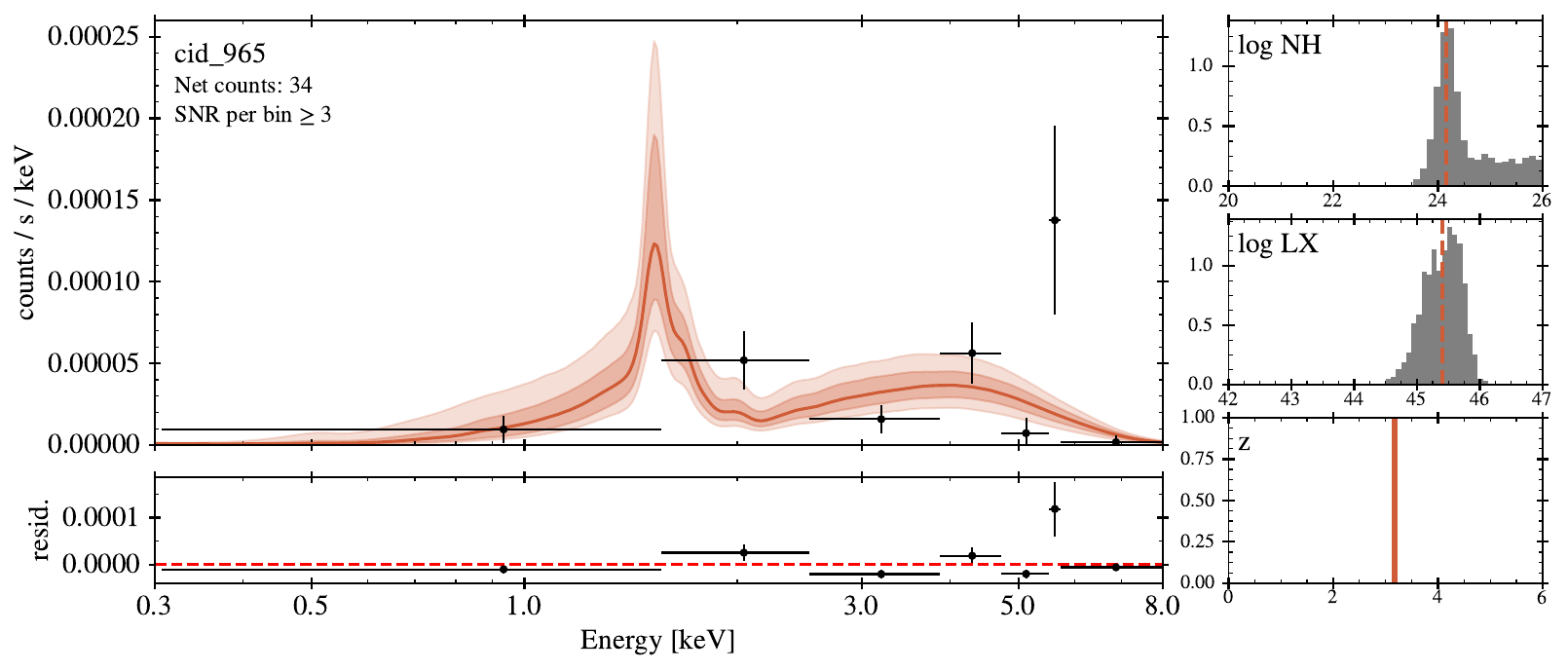}
    \label{fig:xrdata_cid965}
\end{subfigure}
~
\begin{subfigure}[t]{0.49\textwidth}
    \centering
    \includegraphics[width=\linewidth, trim={0 0 0 1.5cm},clip]{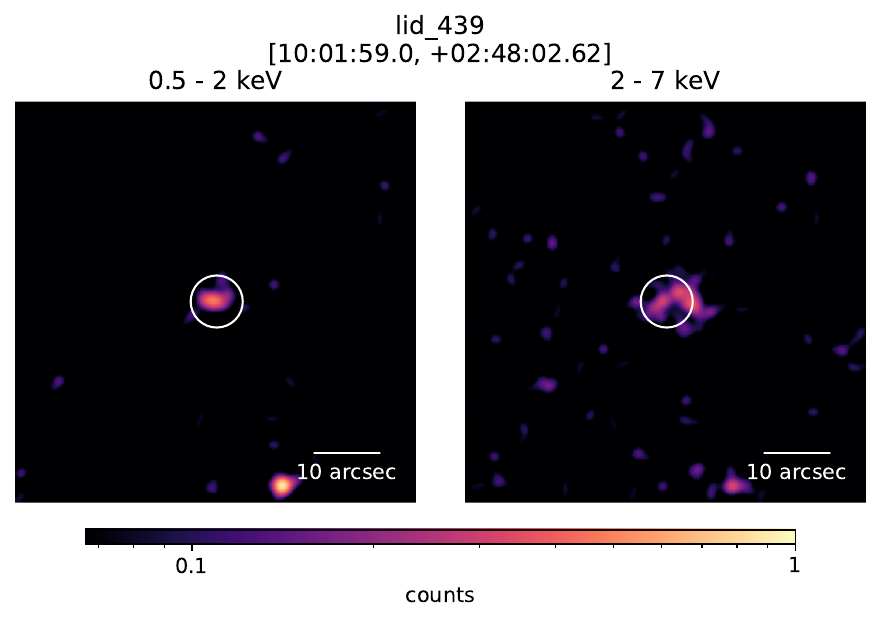}
    \includegraphics[width=\linewidth]{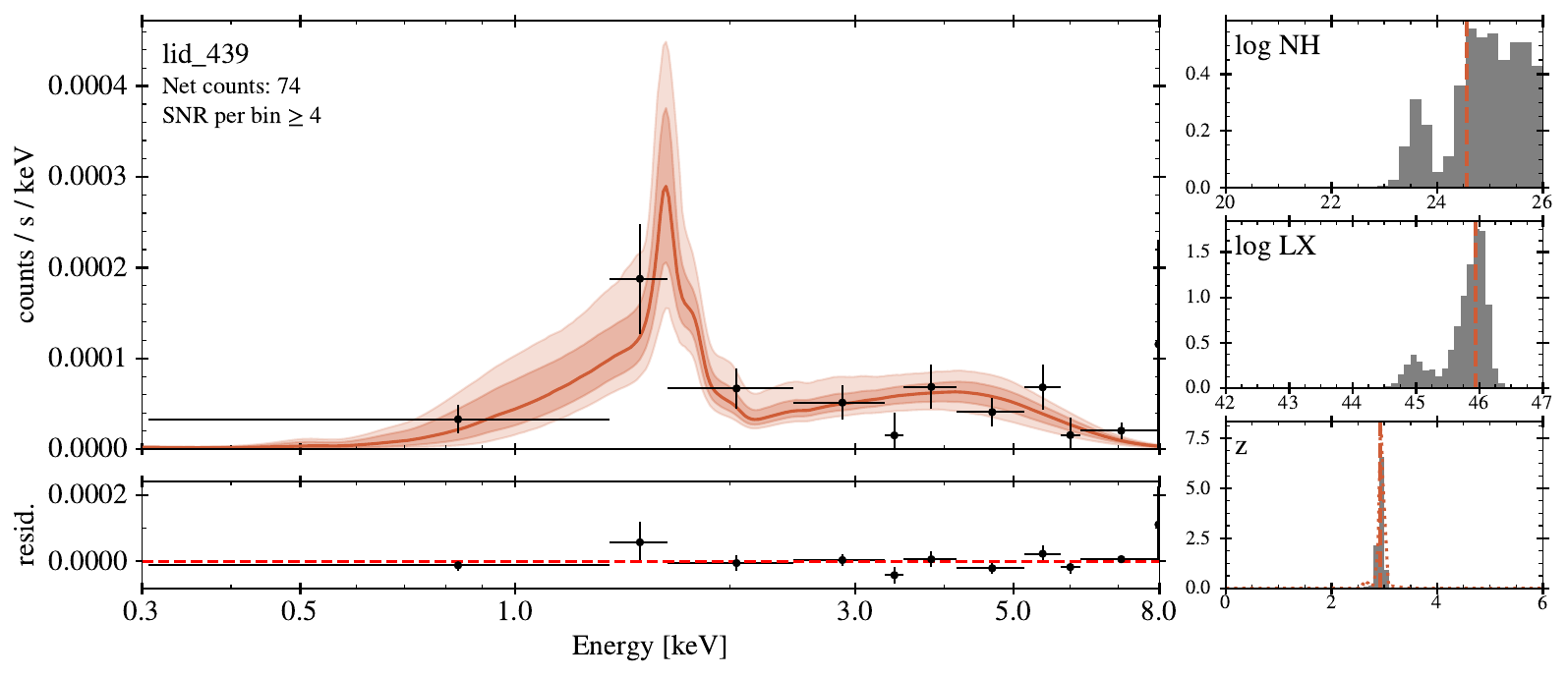}
    \label{fig:xrdata_lid439}
\end{subfigure}

\begin{subfigure}[t]{0.49\textwidth}
    \centering
    \includegraphics[width=\linewidth, trim={0 0 0 1.5cm},clip]{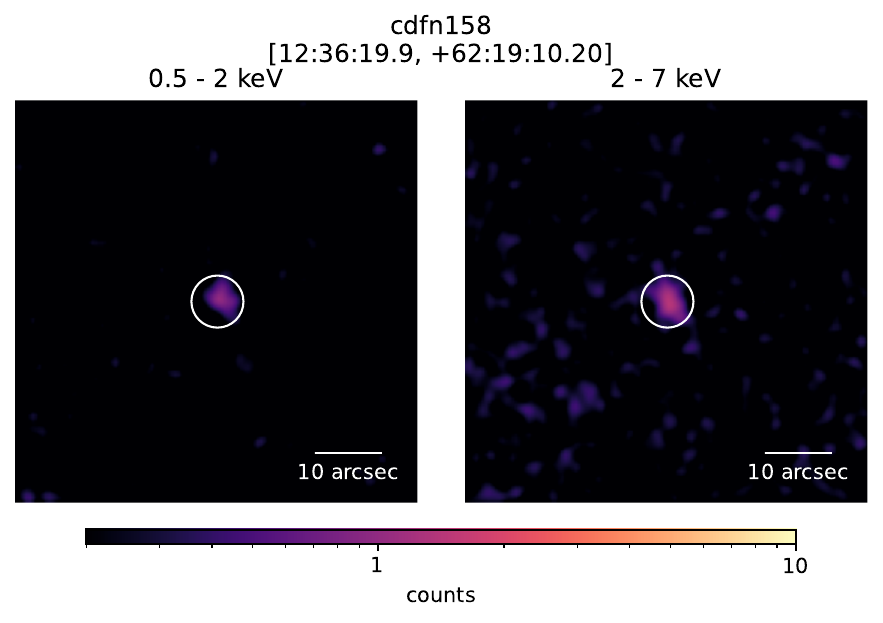}
    \includegraphics[width=\linewidth]{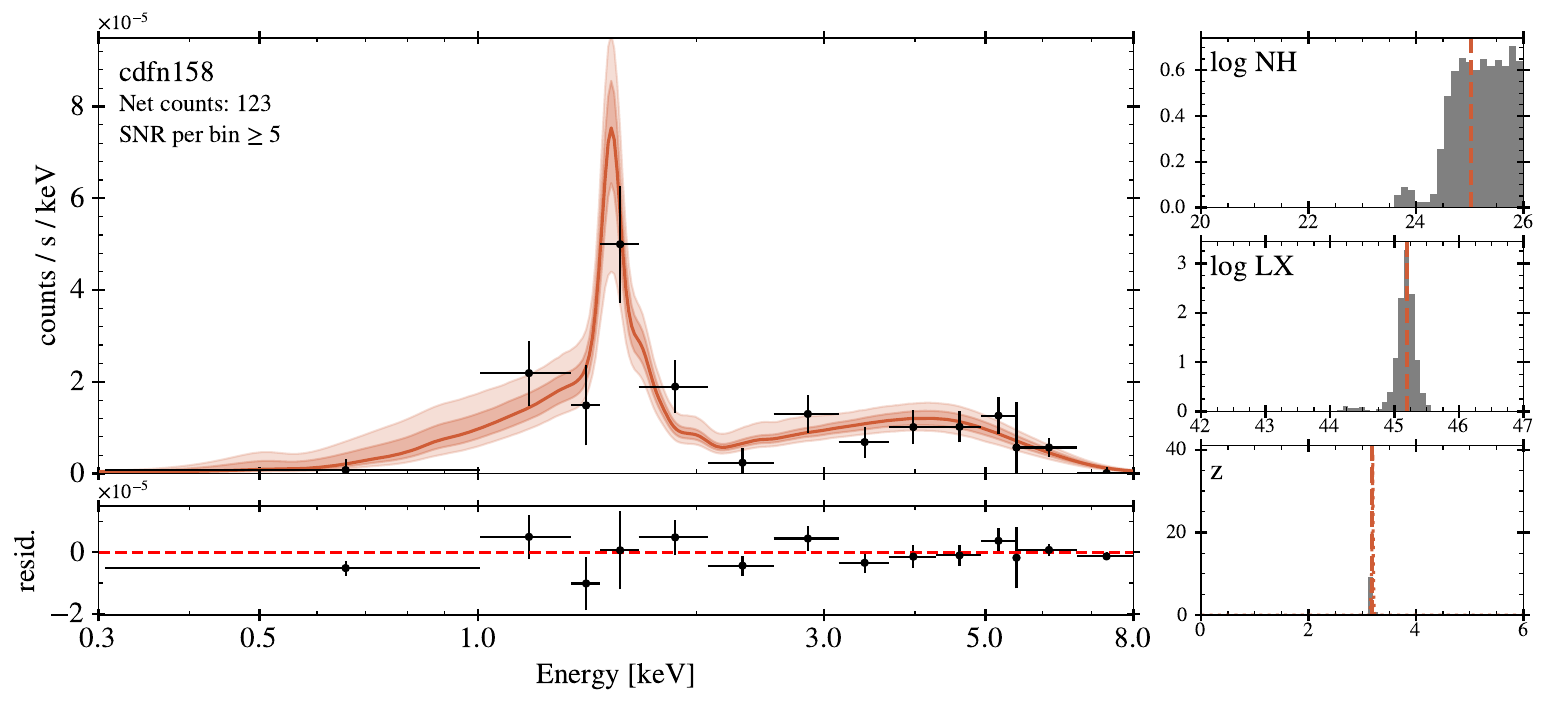}
    \label{fig:xrdata_cdfn158}
\end{subfigure}
~
\begin{subfigure}[t]{0.49\textwidth}
    \centering
    \includegraphics[width=\linewidth, trim={0 0 0 1.5cm},clip]{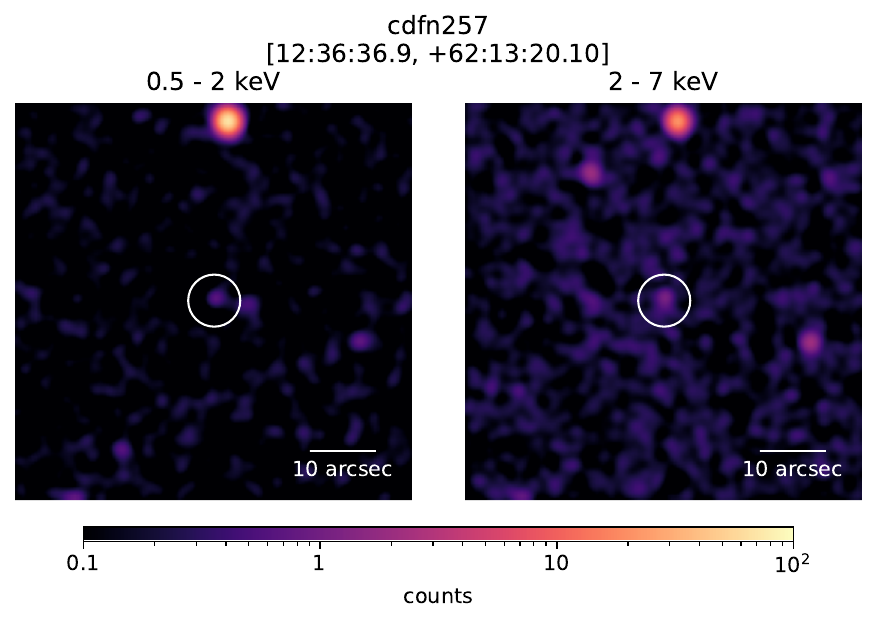}
    \includegraphics[width=\linewidth]{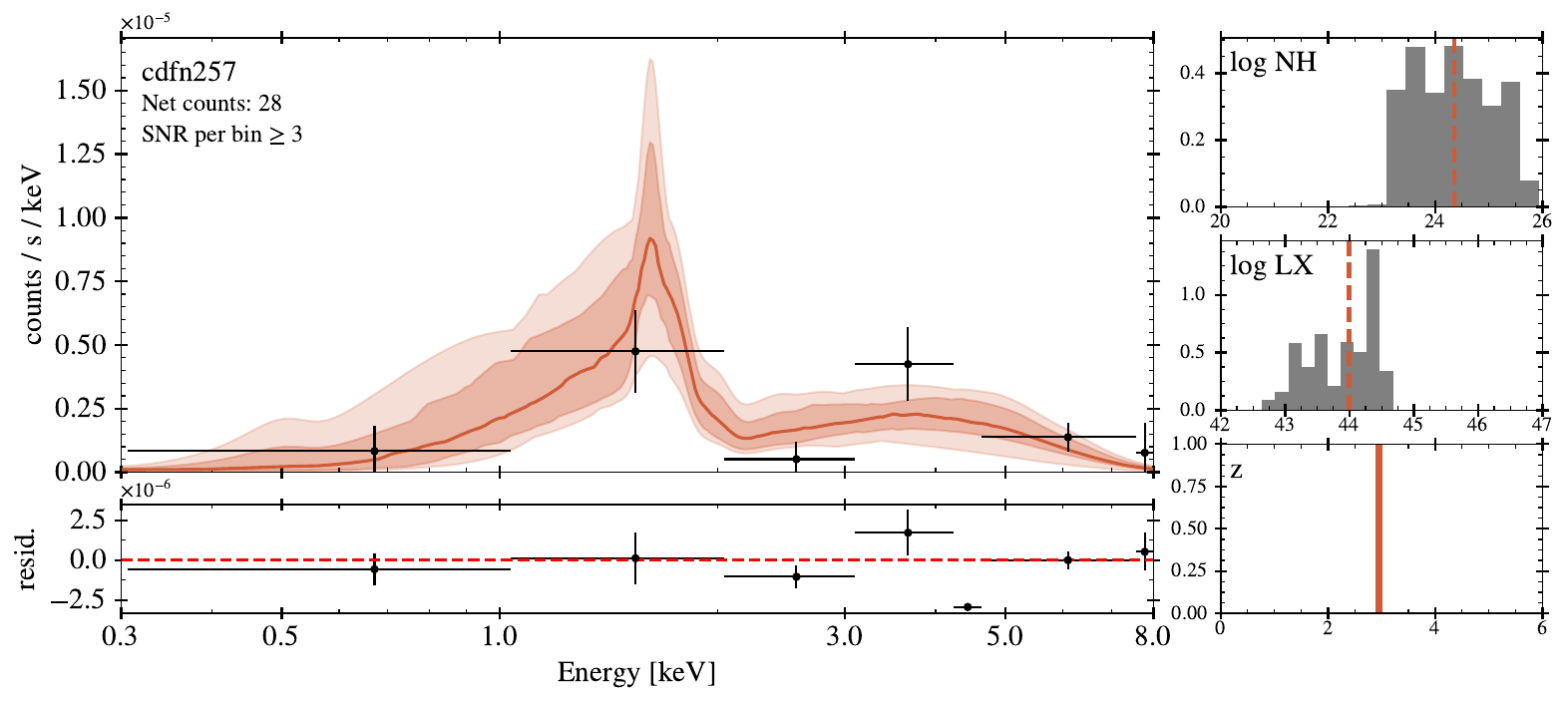}
    \label{fig:xrdata_cdfn257}
\end{subfigure}

\caption{Continued.}
\end{figure*}

\section{Grid of models for CIGALE}
\label{app:cigale}

In this appendix we present the modules and parameter values (see Table~\ref{tab:cigalesettings}) we used in \texttt{CIGALE} for the modelling of the SED of CTK objects.

\begin{table*}
\caption{\texttt{CIGALE} modules and parameter grid.}
\label{tab:cigalesettings}  

\centering
\begin{tabular}{p{0.005\textwidth}lp{0.25\textwidth}p{0.43\textwidth}}
\hline\hline
\multicolumn{3}{l}{Module} & \\  
  & Parameter &  Values & Description \\ 

\hline
\texttt{sfhdelayed} & & & Star formation history: delayed with exponential burst. \\

 & \small \texttt{sfh.age\_main} & 
 [500, 1000, 1500, 2000] Myr & 
 \small Age of the main stellar population. \\
 
 & \small \texttt{sfh.tau\_main} & 
 [200, 500, 700, 1000, 2000] Myr & 
 \small e-folding time of the main stellar population. \\

 & \small \texttt{sfh.age\_burst} & 
 [20] Myr & 
 \small Age of the late burst population. \\

 & \small \texttt{sfh.tau\_burst} & 
 [50] Myr & 
 \small e-folding time of the late burst population. \\

 & \small \texttt{sfh.f\_burst} & 
 [0.0, 0.005, 0.01, 0.015, 0.02, 0.05, 0.10] & 
 \small Mass fraction of the late burst population. \\

\hline
\texttt{bc03} & & & Stellar synthesis population model \citep{Bruzual2003}. \\

 & \small \texttt{stellar.imf} & 
 0 & 
 \small Salpeter's initial mass function \citep{Salpeter1955}. \\

 & \small \texttt{stellar.metallicity} & 
 [0.02] $Z_\sun$ & 
 \small Metallicity of the stellar population. \\

 & \small \texttt{stellar.separation\_age} & 
 [10] Myr & 
 \small Age separation between old and young stellar populations. \\

\hline
\texttt{nebular} & & & Nebular emission. \\

 & \small \texttt{nebular.logU} & 
 [-2.0]  & 
 \small Ionization parameter. \\

 & \small \texttt{nebular.zgas} & 
 [0.014]  & 
 \small Gas metallicity. \\

 & \small \texttt{nebular.ne} & 
 [100] $\mathrm{cm^{-3}}$ & 
 \small Electron density. \\

 & \small \texttt{nebular.f\_esc} & 
 [0]  & 
 \small Fraction of Lyman continuum escaping the galaxy. \\

 & \small \texttt{nebular.f\_dust} & 
 [0]  & 
 \small Fraction of Lyman continuum absorbed by dust.  \\

 & \small \texttt{nebular.lines\_width} & 
 [100] $\mathrm{km~s^{-1}}$ & 
 \small Line width (FWHM).  \\

\hline
\texttt{dustatt\_modified\_CF00} & & & Dust attenuation: modified attenuation law \citep{Charlot2000}. \\
 
 & \small \texttt{attenuation.Av\_ISM} & 
 [0.1, 0.2, 0.3, 0.4, 0.5, 0.6, 0.7, 0.8, 0.9, 1, 1.5, 2, 2.5, 3, 3.5, 4] mag & 
 \small V-band attenuation in the ISM. \\

 & \small \texttt{attenuation.mu} & 
 [0.44] & 
 \small Attenuation ratio between the ISM and the birth clouds. \\

 & \small \texttt{attenuation.slope\_ISM} & 
 [-0.7] & 
 \small Power law slope of the attenuation in the ISM. \\

 & \small \texttt{attenuation.slope\_BC} & 
 [-1.3] & 
 \small Power law slope of the attenuation in the birth clouds. \\

\hline
\texttt{dale2014} & & & Dust emission from the galaxy. \\

 & \small \texttt{dust.alpha} & 
 [1.0, 2.0, 3.0] & 
 \small Slope for the dust emission law. \\

 & \small \texttt{dust.fracAGN} & 
 [0] & 
 \small Fraction of AGN luminosity set to zero. AGN emission is introduced by the SKIRTOR model. \\
 
\hline
\texttt{skirtor2016} & & & AGN model. \\
 
 & \small \texttt{agn.t} & 
 [7, 9, 11] & 
 \small Average edge-on optical depth at $9.7~\microns$. \\

 & \small \texttt{agn.pl} & 
 [1.0] & 
 \small Power-law exponent that sets the radial gradient of dust density. \\

 & \small \texttt{agn.q} & 
 [1.0] & 
 \small Index of the dust density gradient with polar angle. \\

 & \small \texttt{agn.oa} & 
 [40] deg & 
 \small Angle between the equatorial plane and the edge of the torus. \\

 & \small \texttt{agn.R} & 
 [20] & 
 \small Ratio of outer to inner radius of the torus ($R_\mathrm{out}/R_\mathrm{in}$). \\

 & \small \texttt{agn.Mcl} & 
 [0.97] & 
 \small Fraction of total dust mass inside clumps. $0.97$ means $97\%$ of total mass is inside the clumps and $3\%$ in the interclump dust. \\

 & \small \texttt{agn.i} & 
 [30, 50, 70, 90] deg & 
 \small Viewing angle: i = [0, 90 - oa] (face-on, type 1 AGN); i = [90 - oa, 90] (edge-on, type 2 AGN). \\

 & \small \texttt{agn.disk\_type} & 
 1 & 
 \small Disc spectrum from \citet{Schartmann2005}. \\

 & \small \texttt{agn.delta} & 
 [-0.36] & 
 \small Power-law of index $\delta$ modifying the optical slop of the disc. Negative values make the slope steeper where as positive values make it shallower. \\

 & \small \texttt{agn.fracAGN} & 
 [0.0, 0.01, 0.1, 0.2, 0.3, 0.4, 0.5, 0.6, 0.7, 0.8, 0.9, 0.99] & 
 \small Fraction of AGN IR luminosity to total IR luminosity ($1-1000\,\microns$). \\

 & \small \texttt{agn.law} & 
 0 & 
 \small SMC extinction law for the polar dust. \\

 & \small \texttt{agn.EBV} & 
 [0.0, 0.2, 0.4] mag & 
 \small Extinction, E(B-V), in polar direction. \\

 & \small \texttt{agn.emissivity} & 
 [1.6] & 
 \small Emissivity index of the polar dust.  \\
 
 & \small \texttt{agn.temperature} & 
 [100] K & 
 \small Temperature of the polar dust.  \\
\hline
\end{tabular}
\end{table*}

\end{appendix}

\end{document}